\documentclass[usenatbib]{mn2e}

\usepackage{times}
\usepackage{graphicx}
\usepackage{amssymb}
\usepackage{amsmath}

\title[Chaos around black holes with discs or rings]
      {Free motion around black holes with discs or rings:\\
       between integrability and chaos -- IV}
\author[V. Witzany, O. Semer\'ak, P. Sukov\'a]
       {V. Witzany,$^1$\thanks{E-mail: jovter@gmail.com}
        O. Semer\'ak$^1$\thanks{E-mail: oldrich.semerak@mff.cuni.cz}
        and
        P. Sukov\'a$^2$\thanks{E-mail: lvicekps@seznam.cz}\\
       $^1$Institute of Theoretical Physics,
           Faculty of Mathematics and Physics,
           Charles University in Prague,
           Czech Republic\\
       $^2$Institute for Theoretical Physics,
           Polish Academy of Sciences, Warsaw,
           Poland}
\begin{document}

\date{}

\pagerange{\pageref{firstpage}--\pageref{lastpage}} \pubyear{}

\maketitle

\label{firstpage}

\begin{abstract}
The dynamical system studied in previous papers of this series, namely a bound time-like geodesic motion in the exact static and axially symmetric space-time of an (originally) Schwarzschild black hole surrounded by a thin disc or ring, is considered to test whether the often employed ``pseudo-Newtonian" approach (resorting to Newtonian dynamics in gravitational potentials modified to mimic the black-hole field) can reproduce phase-space properties observed in the relativistic treatment. By plotting Poincar\'e surfaces of section and using two recurrence methods for similar situations as in the relativistic case, we find similar tendencies in the evolution of the phase portrait with parameters (mainly with mass of the disc/ring and with energy of the orbiters), namely those characteristic to weakly non-integrable systems. More specifically, this is true for the Paczy\'nski--Wiita and a newly suggested logarithmic potential, whereas the Nowak--Wagoner potential leads to a different picture. The potentials and the exact relativistic system clearly differ in delimitation of the phase-space domain accessible to a given set of particles, though this mainly affects the chaotic sea whereas not so much the occurrence and succession of discrete dynamical features (resonances). In the pseudo-Newtonian systems, the particular dynamical features generally occur for slightly smaller values of the perturbation parameters than in the relativistic system, so one may say that the pseudo-Newtonian systems are slightly more prone to instability.
We also add remarks on numerics (a different code is used than in previous papers), on the resemblance of dependence of the dynamics on perturbing mass and on orbital energy, on the difference between the Newtonian and relativistic Bach--Weyl rings, and on the relation between Poincar\'e sections and orbital shapes within the meridional plane.
\end{abstract}


\section{Introduction}

Newton's theory of gravity is still being used in treating many astrophysical systems, because general relativity is (i) often not necessary in weak-field problems, while (ii) often practically inapplicable (or only applicable numerically) in strong-field ones. Under both circumstances, various approximation methods have been developed, including, above others, ``linearized theory of gravity" and post-Newtonian or post-Minkowskian expansions, as well as ad hoc effective descriptions like those based on ``pseudo-Newtonian" potentials. The well-justified small-parameter expansions are typically reliable in weak-field cases, but in strong field they are inaccurate unless brought to higher expansion orders. The ad hoc formulas, though not derived by any sound approximation scheme, may be quite simple yet still work well in strong field, but much caution is in place, because they often mimic {\em certain} features of the problem accurately, while badly misrepresenting the others. Depending on particular approach, it may be difficult to specify which kinds of errors and of what sizes it brings, the more so if one does not know the stability properties of the exact general relativistic solution or if such a solution is not even available at all.

One of thorough ways to assess the practical quality of a given description of a given gravitational field, or at least its general difference from another description, is to study the motion of free test particles by methods used in the theory of dynamical systems. Though it is problematic to directly compare trajectories of {\em different} dynamical systems and hence to quantify their relative deviation, it is still possible to compare the systems' overall ``dynamical portraits" and the latter's dependence on parameters. Needless to say, the same methods can reveal the effect of various physical perturbations imposed on a given system within {\em the same} theory or approximation; similarly, they can also be employed to test and compare numerical codes.

In the previous three papers of this series \citep{SemerakS-10,SemerakS-12,SukovaS-13} (below referred to as papers I, II and III, respectively), we studied the field of a Schwarzschild-like black hole surrounded by a concentric thin disc or ring, as described by exact static and axially symmetric solution of Einstein equations. Motivated by astrophysical black holes surrounded by accretion (or galactic ``circumnuclear") structures, we analysed the gravitational influence of the additional matter on a long-term dynamics of time-like geodesics and showed, by several different methods, that it can make the dynamics chaotic. In the present paper we compare the previous relativistic results with a similar analysis carried out within pseudo-Newtonian description. More specifically, we emulate the Schwarzschild gravitational field by several simple ``pseudo-Newtonian" potentials, while the disc or ring are described by their Newtonian potential (which equals the first of the two metric functions appearing in the relativistic description).

Besides describing the gravitational field and the free test-particle motion in a different way, we also use a different numerical code to follow the trajectories: whereas the relativistic geodesic-equation system was solved, in previous papers, by the Runge--Kutta (or rather the Hut$\!$'a) 6th-order method with variable proper-time step, here we solve the Newtonian equations of motion by appropriate geometric integrators (see \citealt{HairerWL-06}), specifically in the thin-disc case we have developed an integrator inspired by \cite{SeyrichLG-12} and endowed with a special treatment of the field jump across the disc. In spite of these significant differences, we have arrived at a similar dynamical picture, which justifies the observations made in either of the ways. However, there still occur differences with respect to the exact Schwarzschild picture, and mainly between the different pseudo-Newtonian potentials; some of the latter even do not seem to be reasonably applicable.

After a short note on previous results that have appeared in the literature, we specify the pseudo-Newtonian description  of our gravitational fields in section \ref{pseudo-description} and review basic properties of motion in their backgrounds in section \ref{pseudo-motion}. Then in section \ref{numerics} we give a basic information about the codes employed. The main section \ref{pseudo-exact-comparison} brings the comparison between exact relativistic and pseudo-Newtonian results, using Poincar\'e surfaces of section and two recurrence methods. We add there special notes i) on the link between the dependence on perturbing mass and on orbital energy; ii) on a different character of the relativistic Bach--Weyl ring and of its Newtonian counter-part; and iii) we also point out (and illustrate) that the Poincar\'e sections represent only partial information about the orbits. Concluding remarks then close the paper.

\subsection{Previous results from the literature}

A similar system we consider here was studied by \cite{VokrouhlickyK-98} within Newtonian description and with motivation stemming from a long-term evolution of stars orbiting the black holes (with accretion discs) in galactic nuclei. The authors represented the central body by the $-M/r$ potential and the thin disc by the Kuzmin potential $-{\cal M}/\sqrt{\rho^2+(A+|z|)^2}$, denoting by ${\cal M}$ the disc mass, by $\rho$ and $z$ cylindrical coordinates and by $A$ a free constant, while also taking into account {\em mechanical} effect of the disc on the test orbiter (hydrodynamical drag). The main conclusion was that ``any consistent model of the star-disc interaction has to take the influence of the disc gravity into account, in addition to the effects of direct collisions with gaseous material". The long-term dynamics was found to be sensitive to a particular model of the disc, especially to the radial profile of its surface density, whereas much less to the total mass of the disc.

The {\em pseudo}-Newtonian potentials have been employed in many papers on accretion flows around black holes, but only a few times in studying the chaotic regimes of motion in perturbed black-hole fields.
\cite{GueronL-01} compared the free-motion dynamics around a Schwarzschild black hole and around a Newtonian point centre, when superposed with a dipolar field. They observed that the black-hole system became more chaotic (than the exact case) when the centre was simulated by the Paczy\'nski--Wiita pseudo-potential, mainly if incorporating special relativistic equation of motion.
\cite{SelaruMCG-05} studied the Newtonian circular Hill's restricted three-body problem while describing the primary by the Schwarzschild-type potential $A/r+B/r^3$.
Similarly, \cite{SteklainL-06} compared the Hill problem involving the Paczy\'nski--Wiita pseudo-potential with the original Newtonian version, concluding that the pseudo-Newtonian case is usually -- but not always -- more unstable than its Newtonian counter-part.

Several papers have also tried to incorporate, within the pseudo-Newtonian approach, dragging effects due to rotation of the centre.
\cite{SteklainL-09} thus found that the orbits counter-rotating with respect to the centre are more unstable than the co-rotating ones.
\cite{WangW-11} superposed a rotating ``pseudo black hole" with a quadrupole halo in order to analyse the emission of gravitational waves from orbiting particles; the radiated amplitude and power were observed to be closely related to the degree of orbital chaoticity.
The same authors \citep{WangW-12} also used their model in order to discuss how the geodesic dynamics responds to the centre's spin and to quadrupole perturbation; they found, in particular, that the centre's rotation rather attenuates the instability.
The dynamics of charged particles in the field of a magnetized compact object described in a pseudo-Newtonian manner was then studied by \cite{WangWS-13} and instabilities were identified using the ``fast Lyapunov indicator".

The advance to the pseudo-Newtonian imitation of {\em spinning} fields mainly followed the proposal by \cite{ArtemovaBN-96} of two simple potentials for the Kerr black hole. Recently, these have been checked against a slightly different formula (as well as against the ``benchmark" of the Paczy\'nski--Wiita potential) on the behaviour of circular-orbit acceleration by \cite{KarasA-15}. A more elaborate pseudo-Newtonian ``fit" of Kerr was presented by \cite{ChakrabartiM-06}. \cite{IvanovP-05} found a pseudo-potential for circular motion of a weakly charged particle in the Kerr--Newman space-time. Another extension was suggested by \cite{StuchlikK-08} who derived a generalization of the Paczy\'nski--Wiita prescription for the Schwarzschild--de Sitter black hole.

In order to properly involve rotational dragging, velocity-dependent potentials have also been considered.
\cite{SemerakK-99} tested one such idea against the exact solution on long-term behaviour of the difference between the respective free-particle dynamics. Recently \cite{GhoshSB-14} suggested a new pseudo-potential which reasonably reproduces the Kerr space-time features for moderate centre's angular momentum and moderate energy of the orbiter (see also the overview given in Introduction of that paper, including previous results of its authors). But even in the Schwarzschild case the difference between Newtonian and relativistic dynamics suggests the usage of velocity-dependent expressions; in a thorough study of the pseudo-Newtonian descriptions of the Schwarzschild field, \cite{TejedaR-13} brought such a more advanced possibility (see \citealt{TejedaR-14} for its further development).

\begin{figure*}
\includegraphics[width=0.9\textwidth]{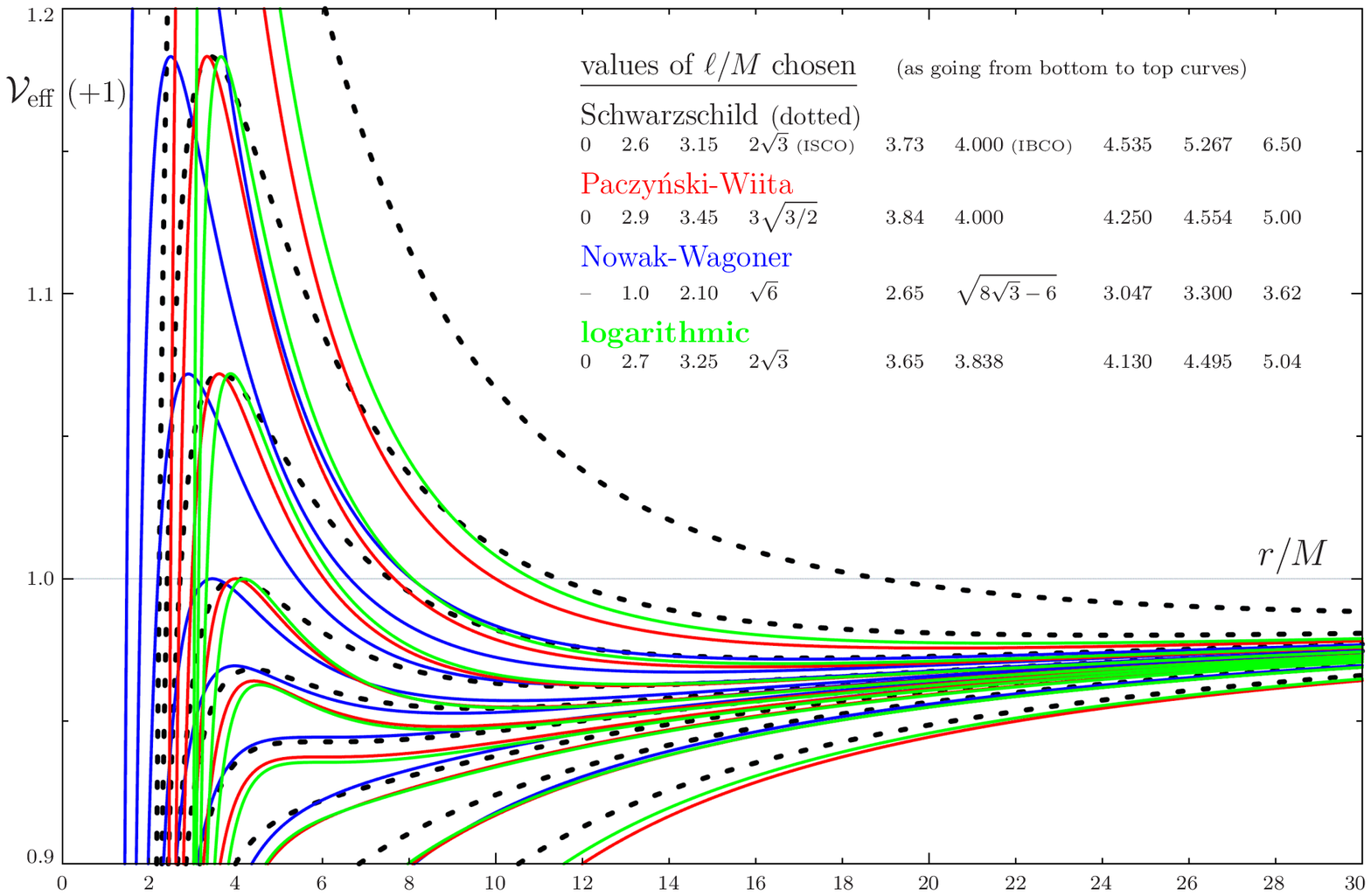}
\caption
{Comparison of effective potentials resulting from the pseudo-Newtonian gravitational potentials (\ref{V_PW}), (\ref{V_NW}) and (\ref{V_ln}) (enlarged by one) with the exact Schwarzschild effective potential $\sqrt{(1-2M/r)(1+\ell^2/r^2)}\,$. Several profiles with different $\ell$ are plotted, with the values of $\ell$ adjusted (differently for different potentials) so that the curves be similar; particular curves involving the marginally stable and marginally bound circular orbits are shown for all the potentials and are easily recognizable.
(For the NW potential, the $\ell=0$ case does not differ much from $\ell/M=1$, so it is not included.)
All the three potentials look similar and not far from the actual Schwarzschild one. Our logarithmic potential is clearly very close to the PW one, having its maxima at slightly larger radii; the NW-potential profiles, on the contrary, are shifted to smaller radii with respect to the PW ones. The main difference is that the slopes of Schwarzschild curves are less steep. Also notice the differences in the values of $\ell$ corresponding to roughly same heights of the potentials: i) those required for the NW potential are considerably lower; ii) at the high-$\ell$ end, those required by the Schwarzschild potential grow faster.}
\label{Veff-comparison}
\end{figure*}

\section{Black hole with disc or ring: a pseudo-Newtonian description}
\label{pseudo-description}

Exact superpositions of a vacuum static axisymmetric (originally Schwarzschild) black hole with a concentric thin disc or ring are described by formulas which were given in the previous papers of this series (see mainly section 1.1 of the last paper III for a compact summary), so rather than repeating them again, let us only specify that we will again choose the inverted 1st member of the counter-rotating Morgan--Morgan thin-disc family (iMM1 disc) and the Bach--Weyl linear ring (BW ring) as the external sources, approximating a thin accretion disc or ring, respectively. Let us also remind that $(t,\rho,z,\phi)$ stand for the Weyl coordinates and $(t,r,\theta,\phi)$ for the Schwarzschild-type coordinates, with $t$ and $\phi$ being Killing time and azimuth and $\rho$, $z$ or $r$, $\theta$ covering the meridional two-space. Geometrized units are used in which $c=G=1$, cosmological constant is (necessarily) set to zero and index-posed commas mean partial differentiation.

Newtonian analogue of the relativistic black-hole--disc/ring picture studied in previous papers is given by function $\nu$ which determines the $g_{tt}$ metric component, in Weyl coordinates satisfies the Laplace equation and represents a direct counter-part of Newton's gravitational potential. We will thus use the metric functions $\nu_{\rm iMM1}$ and $\nu_{\rm BW}$ of the disc and of the ring directly as the disc or ring Newtonian potentials, respectively. The Schwarzschild-centre potential $\nu_{\rm Schw}$, on the other hand, will be just emulated by a certain effective pseudo-potential. We will test three simple cases,
\begin{align}
  V_{\rm PW} &= -\frac{M}{r-2M} \;,
                \label{V_PW} \\
  V_{\rm NW} &= -\frac{M}{r}\left(1-\frac{3M}{r}+\frac{12M^2}{r^2}\right),
                \label{V_NW} \\
  V_{\rm ln} &= \frac{1}{3}\,\ln\left(1-\frac{3M}{r}\right).
                \label{V_ln}
\end{align}
The first was proposed by \cite{PaczynskiW-80}, the second by \cite{NowakW-91}, and the logarithmic one represents another possibility we are submitting for comparison.
The Paczy\'nski-Wiita potential is a default benchmark, very simple yet behaving surprisingly well in many situations. The logarithmic potential is simply included because we newly suggest it here. And the Nowak-Wagoner potential is chosen for it has yet another form which will be seen to result in a rather different character of the accessible phase-space region; at the same time, it has turned out to be the best of ``simple" possibilities in some studies (e.g. \citealt{Crispino-etal-11}).

\begin{figure}
\includegraphics[width=\columnwidth]{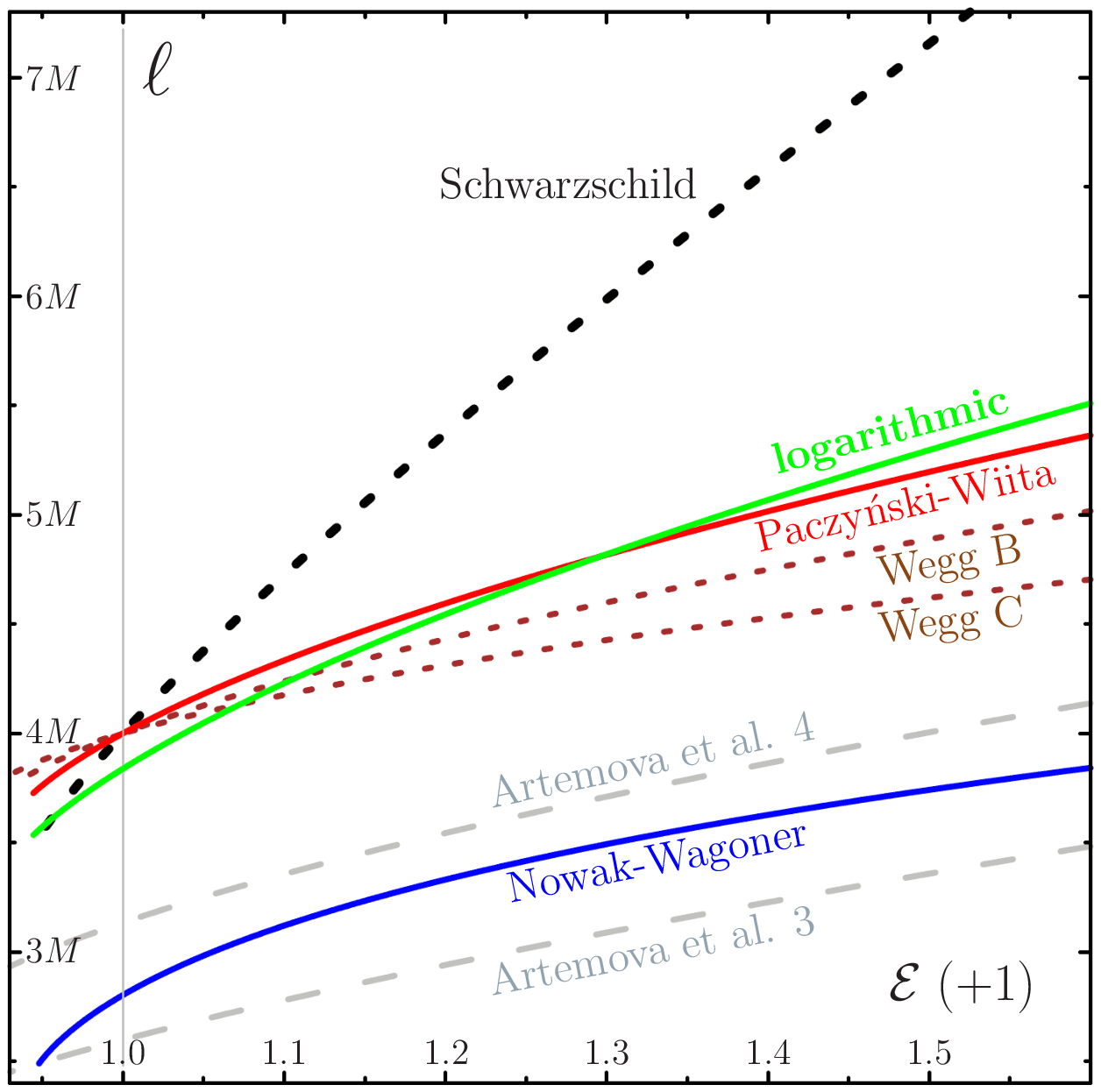}
\caption
{Values of the angular momentum $\ell$ needed to raise the centrifugal barrier to a given energetic level ${\cal E}$ (thus to establish an unstable circular orbit with that energy), plotted for the potentials we compare (${\cal E}$ is enlarged by 1 for the potentials in order to match the relativistic case). The Nowak--Wagoner potential yields the worst result and our logarithmic potential yields the best one, yet none of them reproduces the Schwarzschild-field behaviour properly. The curves provided by potentials (\ref{V_ABN3}) and (\ref{V_ABN4}) of Artemova et al. (1996) are also shown in dashed grey and the potentials (\ref{V_WB}) and (\ref{V_WC}) of Wegg (2012) are drawn in dotted brown.}
\label{ell-on-En}
\end{figure}

Other major simple pseudo-Newtonian substitutes for Schwarzschild were provided by \cite{ArtemovaBN-96} and quite recently by \cite{Wegg-12}. \cite{ArtemovaBN-96} used several pseudo-potentials in studying  disc accretion onto black holes; in the non-rotating case, they considered expressions (we number them according to the original paper)
\begin{align}
  V_{\rm ABN3} &= -1+\sqrt{1-\frac{2M}{r}} \;,  \label{V_ABN3} \\
  V_{\rm ABN4} &= \frac{1}{2}\;\ln\left(1-\frac{2M}{r}\right). \label{V_ABN4}
\end{align}
The second one (just equal to the Schwarzschild potential $\nu_{\rm Schw}$) is similar in form to our logarithmic expression (\ref{V_ln}), but we will see that the latter is actually more similar to the PW potential (see Figs. \ref{ell-on-En} and \ref{eff-potentials-1valley}). A comparison of the two ABN potentials with the PW and NW ones was performed by \cite{Crispino-etal-11} on the motion of a particle emitting scalar radiation.
More recently, several serious options have been presented by \cite{Wegg-12} (original marking by A, B and C is kept again),
\begin{align}
  V_{\rm WA} &= -\frac{M}{r}\left(1+\frac{3M}{r}\right),  \label{V_WA} \\
  V_{\rm WB} &= -\frac{M}{r}\left(\frac{3r}{3r-5M}+\frac{4M}{3r}\right),  \label{V_WB} \\
  V_{\rm WC} &= -\frac{M}{r}\,
                 \frac{1+\frac{4M}{r}(3-\sqrt{6})+\frac{20M^2}{r^2}(5-2\sqrt{6})}
                      {1-\frac{M}{r}(4\sqrt{6}-9)} \;,  \label{V_WC}
\end{align}
and shown to yield better results for the apsidal precession of low-energy (about parabolic) orbits than the Paczy\'nski-Wiita potential. Recently we have included, with a surprisingly good result, $V_{\rm WA}$ in a comparison of light-ray approximations in the Schwarzschild field \citep{Semerak-15}. However, this potential turns out to be unsuited for our present purposes as shown in the next section (equation (\ref{Wegg,effective}) and below).
All the other four potentials, $V_{\rm ABN3}$, $V_{\rm ABN4}$, $V_{\rm WB}$ and $V_{\rm WC}$, are included in Figs. \ref{ell-on-En} and \ref{eff-potentials-1valley} in order to at least illustrate their basic nature against those we are going to study in more detail in the present paper.

\subsection{Issue of comparison in coordinates}

When preparing to superpose the centre-describing potentials with $\nu_{\rm iMM1}$ or $\nu_{\rm BW}$, one encounters the main query, however:
How exactly to perform the Newtonian superposition in order to get a plausible counter-part of the relativistic system? Which coordinates covering the curved relativistic space-time are adequate counter-parts of Euclidean coordinates of the Newtonian description?
The Newtonian pseudo-potential for the black hole is usually given in Euclidean spherical coordinates and simulates the hole represented in Schwarzschild coordinates, while the disc/ring potentials are naturally taken over from relativity in cylindrical coordinates. In the relativistic description, the linear superposition holds in Weyl coordinates $\rho$, $z$ which are of cylindrical type and where the black-hole horizon appears as a finite line singularity at $\rho=0$, $|z|\leq M$. After transformation to Schwarzschild-like coordinates of spheroidal type,
\begin{equation}  \label{rho,z-Weyl}
  \rho=\sqrt{r(r-2M)}\,\sin\theta, \qquad z=(r-M)\cos\theta,
\end{equation}
the black-hole horizon becomes spherical, while the disc/ring keeps its shape but has a slightly bigger coordinate radius.

The spheroidal character of the black hole is clearly not well represented in the Weyl coordinates. However, since the relativistic superposition is performed in them, one should probably reproduce it in the Newtonian approach in the following way:
i) take the pseudo-Newtonian potential (in spherical coordinates) and transform it into the Weyl coordinates;
ii) add the disc or ring potential expressed in the Weyl coordinates;
iii) transform the result to the Schwarzschild-type coordinates.
Since the Newtonian fields superpose linearly in {\em any} coordinates, one can summarize this without the intermediate step: take the black-hole pseudo-potential and add to it the disc or ring potential transformed from cylindrical to spherical/spheroidal coordinates in a Weyl-like manner, i.e. by substituting (\ref{rho,z-Weyl}).

Alternatively, rather than to take over the transformation between the Weyl and Schwarzschild coordinates to the Newtonian description, one could assume that the relativistic disc/ring potential in Weyl coordinates corresponds to the Newtonian one in common cylindrical coordinates, connected with the spherical ones (in which the pseudo-potential for the centre is given) by the Euclidean relation $\rho=r\sin\theta$, $z=r\cos\theta$. However, since the pseudo-potentials should imitate the black hole, which means mainly to imitate the occurrence of the horizon, it is reasonable to demand that the spheroidal-cylindrical transformation have in both cases similar effect on the central source: if it shrinks the relativistic source into a rod, it should not leave the Newtonian source intact (as the Euclidean-type relation). We have anyway tested this second possibility too and learned that if the external source is not very close to the centre (below $10M$, say), the results are almost identical.

However, carefully as one may try to consider the correspondence between the relativistic and pseudo-Newtonian systems, they inevitably remain different, the more so that not only the space(-time) backgrounds differ, but also the dynamics (equations of motion), so one should at least expect a quantitative discrepancy, unless employing some more sophisticated velocity-dependent potential.

\section{Motion in modified Newtonian potentials}
\label{pseudo-motion}

The motion of test particles in the velocity-independent axially symmetric Newtonian potential $V(r,\theta)$ is described, in spherical coordinates $(r,\theta,\phi)$ and with obvious notation, by equations
\begin{align}
  \ddot{r}         &= -V_{,r}+r\,(\dot\theta^2+\dot\phi^2\sin^2\theta),
                   \label{ddr,axi} \\
  r^2\ddot{\theta} &= r^2\dot\phi^2\sin\theta\cos\theta-2r\dot{r}\dot\theta-V_{,\theta} \,,
                   \label{ddtheta,axi} \\
  r^2\ddot{\phi}   &= -2r\dot\phi\,(\dot{r}+r\dot\theta\,\cot\theta).
                   \label{ddphi,axi}
\end{align}
If the field is even {\em spherically} symmetric, $V=V(r)$, the $V_{,\theta}$ term in the 2nd equation vanishes and the motion gets confined to a plane. The orbital plane is usually identified with $\theta=\pi/2$, so one is left with equations
\begin{equation}  \label{ddot:r,phi}
  \ddot{r}=-V_{,r}+r\dot{\phi}^2, \quad
  r\ddot{\phi}=-2\dot{\phi}\,\dot{r} \,.
\end{equation}
These have energy and angular-momentum integrals
\begin{equation}
  E=\frac{m}{2}\,(\dot{r}^2+r^2\dot{\phi}^2)+mV,
  \quad
  L=m r^2\dot{\phi} \,,
\end{equation}
which invert for velocities as
\begin{equation}
  \dot{\phi}=\frac{\ell}{r^2} \,,
  \quad
  \dot{r}^2=\frac{2mr^2(E-mV)-L^2}{m^2 r^2}
           \equiv 2\,({\cal E}-{\cal V}_{\rm eff}) \,,
\end{equation}
where\footnote
{As noted in figures and their captions, we actually shift the specific energy ${\cal E}$ by one so that a particle at rest at infinity has ${\cal E}=1$ in accord with the relativistic case.}
\begin{equation}
  {\cal V}_{\rm eff}:=V+\frac{\ell^2}{2r^2} \,,
  \qquad
  {\cal E}:=\frac{E}{m} \,,
  \quad
  \ell:=\frac{L}{m} \,.
\end{equation}

Circular orbits exist where
\begin{equation}
  {\cal V}_{{\rm eff},r}=0
  \quad\Leftrightarrow\quad
  \ell^2=r^3 V_{,r} \;,
\end{equation}
so their linear speed amounts to
\begin{equation}
  r\dot\phi = \sqrt{rV_{,r}} \;,
\end{equation}
their energy is given by the corresponding potential value
\begin{equation}
  {\cal E}(\ell^2\!=\!r^3 V_{,r})={\cal V}_{\rm eff}(\ell^2\!=\!r^3 V_{,r})
  =V+\frac{1}{2}\,rV_{,r}
\end{equation}
and their stability is determined by the sign of
\begin{equation}
  {\cal V}_{{\rm eff},rr}(\ell^2\!=\!r^3 V_{,r})=V_{,rr}+\frac{3V_{,r}}{r} \,.
\end{equation}

\begin{figure}
\includegraphics[width=\columnwidth]{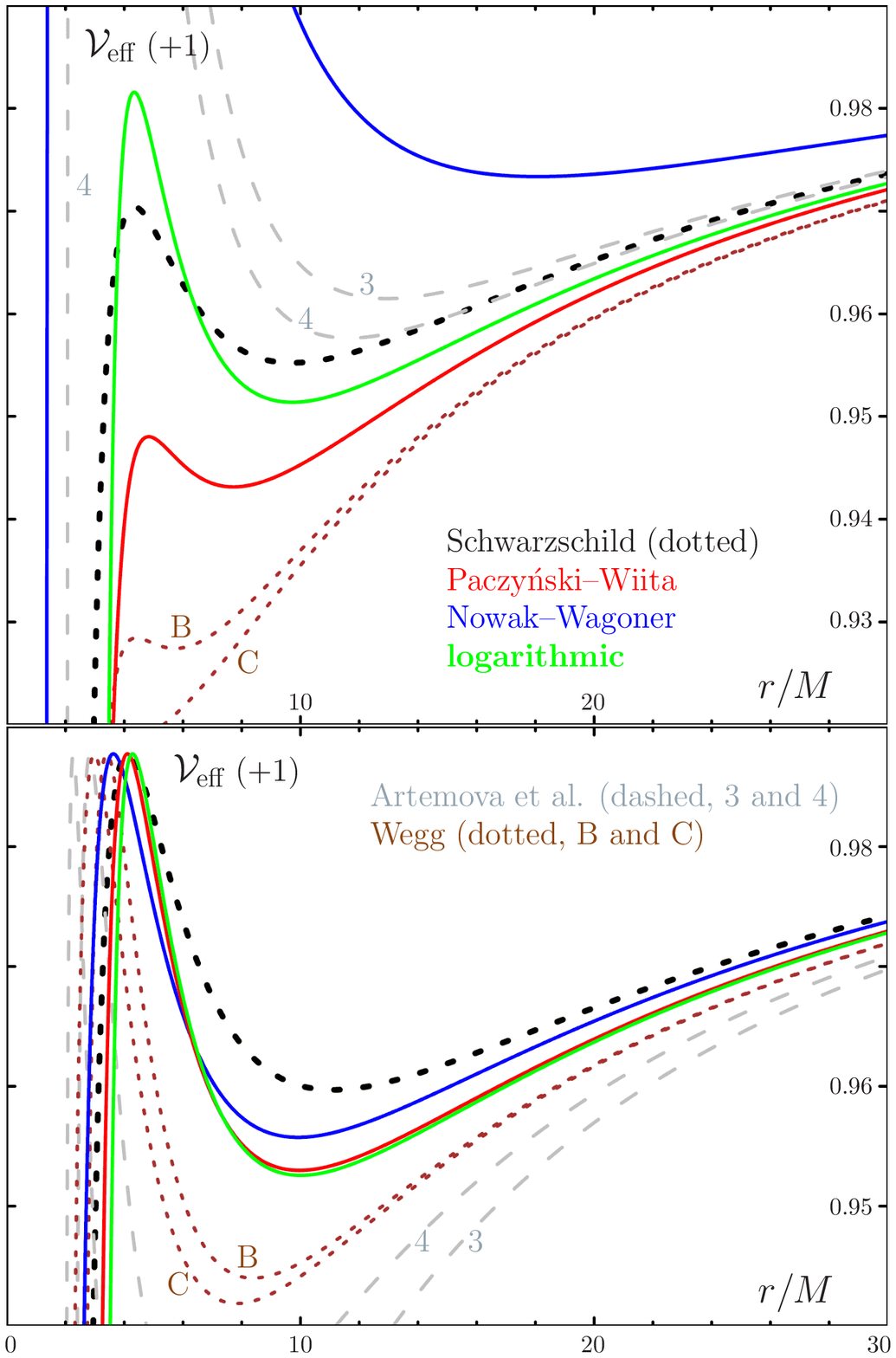}
\caption
{{\it Top:}
One specific effective-potential profile plotted for all the gravitational potentials considered, with the angular momentum $\ell=3.75M$ (this value is chosen in most of the figures presented in next sections). Like in Fig. \ref{ell-on-En}, the Artemova--et-al. potentials $V_{\rm ABN3}$ and $V_{\rm ABN3}$ are also shown in dashed grey and the Wegg potentials $V_{\rm WB}$ and $V_{\rm WC}$ are drawn in dotted brown. Only the PW and the ln potentials (red and green) seem to approximate the exact Schwarzschild shape in some way. Clearly the PW potential is more open towards the centre, while the ln potential is more closed than the actual Schwarzschild case.
{\it Bottom:}
Similar plot, but with the angular-momentum values chosen so that all the effective potentials have the same maximum ${\cal E}+1=0.987746$ at the unstable circular orbit (in the Schwarzschild case, one takes just ${\cal E}$). Concretely, this means $\ell=3.9M$ for Schwarzschild, $\ell=3.9494M$ for Paczy\'nski--Wiita, $\ell=2.7475M$ for Nowak--Wagoner, $\ell=3.7805M$ for the logarithmic potential, $\ell_3=2.5739M$ and $\ell_4=3.1028M$ for the Artemova--et-al. potentials $V_{\rm ABN3}$ and $V_{\rm ABN3}$, and $\ell_{\rm B}=3.9651M$ and $\ell_{\rm C}=3.9735M$ for the Wegg potentials $V_{\rm WB}$ and $V_{\rm WC}$. The pseudo-potentials yield somewhat different radii of the unstable circular orbit (only the PW and ln potentials have it very close to the correct value) and their valley existing above this orbit is deeper than the Schwarzschild one; the difference is especially large for the ABN potentials.}
\label{eff-potentials-1valley}
\end{figure}

The character of radial motion and its response to perturbations are thus governed by shape of the potential well (given by $V$ and $\ell$) and by the particle's specific energy ${\cal E}$.
Most importantly, the shape of ${\cal V}_{\rm eff}$ and the value of ${\cal E}$ determine the properties of the region accessible to the particle within the $(r,\dot{r})$ diagram. A well known crucial point is whether this region is closed or open towards the centre, which, for a given energy, depends on the height of the centrifugal barrier $\ell^2/r^2$. In the marginally closed state, the accessible domain is bounded by a separatrix which corresponds to a homoclinic orbit, winding -- in infinite past and infinite future -- from and on the unstable circular periodic orbit residing at the potential saddle-point vertex. Homoclinic orbits, a salient feature of black-hole fields, represent an infinite-whirl limit of the zoom-whirl type of motion (a strong-field bound motion with extreme pericentre advance), and are familiar to mark the frontiers of chaotic regime -- their perturbation leads to the occurrence of a ``homoclinic tangle", through which the original circular orbit breaks up into a fractal set of periodic orbits.

The homoclinic orbit is infinite, but the length of its trail in reasonable coordinates ($r,\dot{r}$ in our case), i.e. of the accessible-region bounding separatrix, indicates the size of a phase-space region which turns chaotic under perturbation. This does not provide any plausible (``covariant") measure of what fraction of the phase space will be affected, but still can be used to compare different potentials. A similar suggestion (only given by $\dot{r}^2$ rather than by $\dot{r}$) is contained in the length of the potential valley ${\cal V}_{\rm eff}(\ell_{\rm circ})$ below the energy level of the unstable circular orbit or in this valley's area.

Let us now briefly check the basic properties of effective potentials given by the gravitational potentials (\ref{V_PW})--(\ref{V_ln}), in particular whether and how they reproduce circular periodic orbits, decisive for the response of the dynamical system to perturbation.
However, consider first the Wegg's expression (\ref{V_WA}) in order to realize why it is not suitable this time. For the corresponding effective potential,
\begin{equation}  \label{Wegg,effective}
  {\cal V}_{\rm eff}=-\frac{M}{r}\left(1+\frac{3M}{r}\right)+\frac{\ell^2}{2r^2} \,,
\end{equation}
the condition for circular orbits $\ell^2=r^3 V_{,r}$ yields $Mr=\ell^2-6M^2$ for the radius, so $\ell^2>6M^2$ must hold in order that such radii really exist. But for the Wegg potential one has ${\cal V}_{{\rm eff},rr}=(\ell^2-6M^2)/r^4$, so all the $\ell^2>6M^2$ circular orbits sit at the potential {\em minimum}, hence they are all {\em stable} and not interesting for us. Therefore, rather than mimicking the occurrence of unstable circular orbits, so characteristic to the black-hole fields, the Wegg's A-potential behaves like Newtonian $-M/r$, just with the critical value of $\ell^2$ shifted from zero to $6M^2$. (This is no wonder, since Wegg suggested the potential specifically for near-parabolic orbits at larger radii.)

The shapes of the effective potentials resulting from the Paczy\'nski--Wiita, Nowak--Wagoner and our logarithmic potentials is compared in Fig. \ref{Veff-comparison}. All the three potentials host both stable and unstable circular orbits and are clearly quite similar. They all yield the correct radius $r=6M$ for the marginally stable circular orbit (ISCO). The Paczy\'nski--Wiita potential also does so for the marginally bound orbit (IBCO, $r=4M$), reproducing besides the angular-momentum Schwarzschild value $\ell=4M$ there. On the other hand, the logarithmic potential gives the correct value of angular momentum at the ISCO ($\ell=2\sqrt{3}\,M$). The latter is a consequence of a more general tuning: circular orbits of the logarithmic potential satisfy
\begin{equation}
  \ell^2 = \frac{Mr^2}{r-3M}
\end{equation}
which is exactly the same expression as would be obtained in the exact Schwarzschild field. This means, in particular, that a Keplerian disc in the ln-potential would have exactly the same distribution of angular momentum as in the Schwarzschild case.

Figure \ref{ell-on-En} emphasizes what may not be evident from Fig. \ref{Veff-comparison}: that although the shapes of the potentials seem similar to the actual Schwarzschild one, they may differ significantly or just fail in some important aspects like the relations between the energy and angular momentum for the unstable circular periodic orbits. Specifically, a particle with ${\cal E}$, $\ell$ located {\em below} the respective curve in Fig. \ref{ell-on-En} will orbit in an allowed region open towards the center and will thus be prone to black hole in-fall; on the other hand, particles from {\em above} the curve will orbit in two distinct regions, the exterior one being closed-off from the center by the centrifugal barrier. However, if one picks ${\cal E}(+1)<1$ (hoping for bound motion later harbouring chaos) and $\ell$ too far above the curve, there might be {\em no} bound particles orbiting the black hole because of a too high centrifugal barrier. Hence, in the ${\cal E}(+1)<1$ range the PW, Wegg B and C, and log potentials are expected to exhibit satisfactory behaviour in terms of the overall nature of the allowed region, whereas the NW and Artemova potentials will not show a good correspondence.

One can judge from this that although the character of chaos induced by perturbation of the pseudo-Schwarzschild fields is likely to be similar to what is a common experience from weakly non-integrable systems, its dependence on the relevant parameters will be quantitatively different, in particular the parameter values critical for an occurrence of various features (resonances, separatrices, chaotic layers) will be different. Also, as the potential valleys provided by the pseudo-potentials are generically deeper than the actual Schwarzschild ones (see Fig. \ref{eff-potentials-1valley}), it might be loosely anticipated that the corresponding Newtonian motion will rather be more chaotic than geodesic motion in the exact relativistic field. However, one must remember that we are yet talking about the central black hole only, and, also, that the relativistic dynamics is different from Newtonian (already special-relativity effects make some difference), so the centre's effective-potential shape is just one part of the story.

\begin{figure*}
\includegraphics[width=\textwidth]{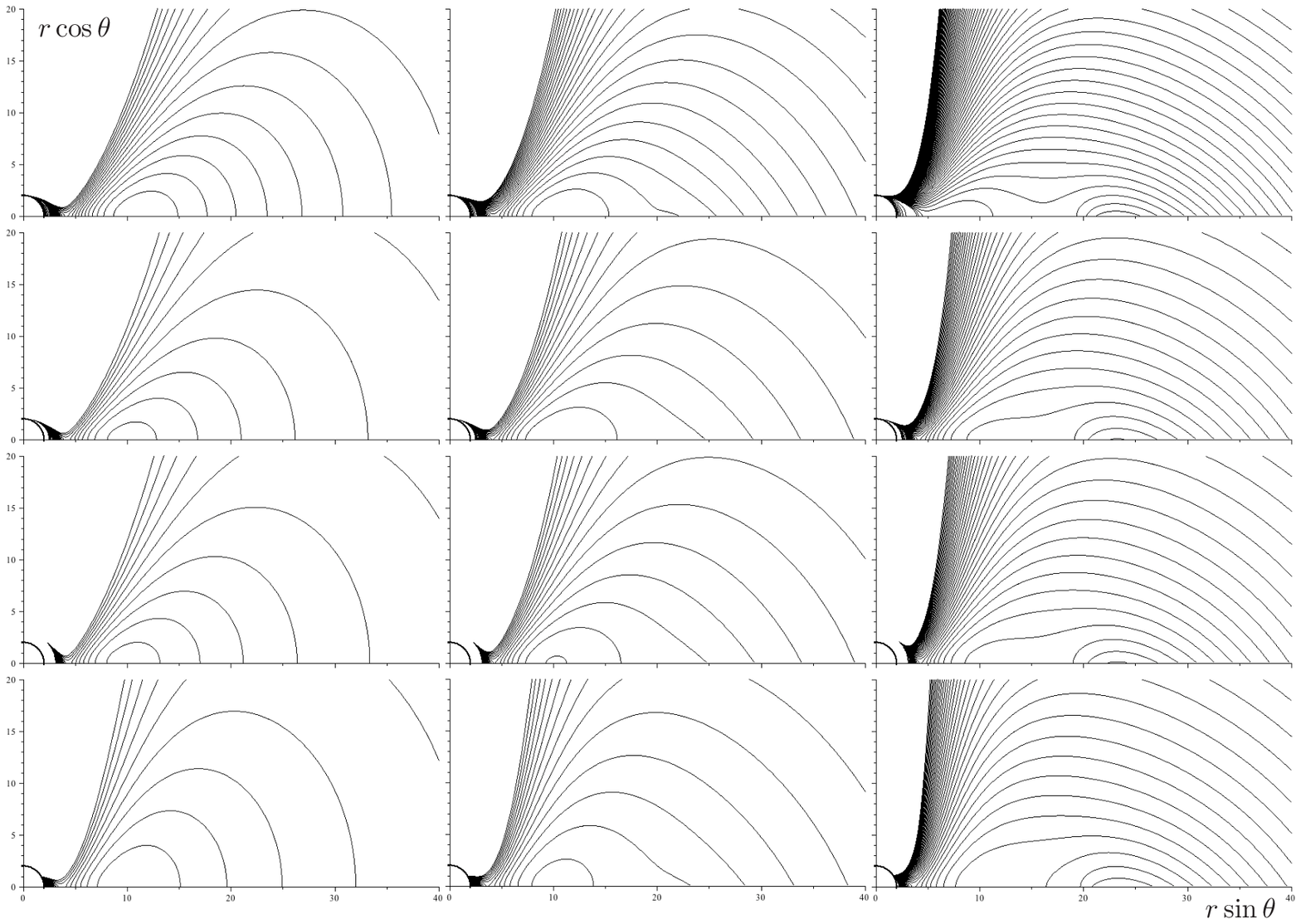}
\caption
{Meridional ($\phi={\rm const}$) sections of the effective potentials for an originally Schwarzschild black hole surrounded by the 1st member of the Morgan--Morgan counter-rotating thin-disc family. Exact relativistic superposition is shown (1st row) together with those involving Paczy\'nski--Wiita (2nd row), logarithmic (3th row) and Nowak--Wagoner (4rd row) imitations of the black hole. All the cases are determined by the value of specific energy ${\cal E}(+1)=0.987746$ at the unstable circular orbit as in Fig. \ref{eff-potentials-1valley} (so the values of $\ell$ are also exactly as there). Within all of the four rows, the disc relative mass ${\cal M}/M$ is chosen, from left to right, $0.0$, $1.0$ and $5.0$. In all the plots, the contours shown are ${\cal V}_{\rm eff}(+1)=0$, $0.1$, $0.2$, $0.3$, \dots $0.700$, $0.705$, $0.710$, $0.715$, \dots, $0.990$, $0.995$, $1$. (${\cal V}_{\rm eff}+1$ is taken in the Newtonian cases, whereas just ${\cal V}_{\rm eff}$ in the Schwarzschild one.) Axes are scaled in the units of $M$.}
\label{BH-iMM1-merid}
\end{figure*}

\subsection{Superposition of the black hole with a disc or ring}

The second part is the gravitational potential of the additional source which in our case will be represented by a thin annular disc or a ring. If a static and axially symmetric source is placed around the centre, the field is no longer spherically symmetric, hence a generic motion is no longer plane-like and one must return to equations (\ref{ddr,axi})--(\ref{ddphi,axi}). Their energy and angular-momentum integrals now have the form
\begin{equation}
  {\cal E}=\frac{1}{2}\,(\dot{r}^2+r^2\dot{\theta}^2+r^2\dot{\phi}^2\sin^2\theta)+V,
  \quad
  \ell=r^2\dot{\phi}\,\sin^2\theta \,,
\end{equation}
and invert for velocities as
\begin{equation}  \label{velocities}
  \dot{\phi}=\frac{\ell}{r^2\sin^2\theta} \,,
  \quad
  \dot{r}^2+r^2\dot\theta^2=2\,({\cal E}-{\cal V}_{\rm eff}) \,,
\end{equation}
where
\begin{equation}
  {\cal V}_{\rm eff}:=V+\frac{\ell^2}{2r^2\sin^2\theta} \,.
\end{equation}

To obtain an effective potential for the motion in the complete field of the (pseudo) black hole surrounded by some external source (which generates potential $\nu_{\rm ext}$), one simply takes the above ${\cal V}_{\rm eff}$ with
\[V=V_{\rm pseudo}(r)+\nu_{\rm ext}\!\left(\!\sqrt{r(r-2M)}\,\sin\theta,(r-M)\cos\theta\right).\]
We illustrate the possible outcome by adding the inverted 1st Morgan--Morgan counter-rotating disc which was already involved in previous papers of this series and whose gravitational potential reads, in the Weyl-type cylindrical coordinates (\ref{rho,z-Weyl}),
\begin{align}
  \nu_{\rm disc}=
 -\frac{\cal M}{\pi(\rho^2+z^2)^{3/2}}
  & \left[\left(2\rho^2\!+\!2z^2\!-\!b^2\frac{\rho^2\!-\!2z^2}{\rho^2+z^2}\right)
          {\rm arccot}\,{\cal S} \right. \nonumber \\
  & \left. \; -\frac{1}{2}\,(3\Sigma-3b^2+\rho^2+z^2)\,{\cal S}\right]
\end{align}
(see e.g. \citealt{ZacekS-02}), where
\begin{align*}
  \Sigma &:= \sqrt{(\rho^2-b^2+z^2)^2+4b^2 z^2} \;, \\
  {\cal S} &:= \sqrt{\frac{\Sigma-(\rho^{2}-b^{2}+z^{2})}{2\,(\rho^{2}+z^{2})}} \;,
\end{align*}
and ${\cal M}$ and $b$ denote mass and Weyl inner radius of the disc.
Figure \ref{BH-iMM1-merid} shows the results obtained with different pseudo-potentials and compares them with the one following from an exact relativistic treatment which describes the problem (geodesic motion) by equations (see section 4 of paper I)
\begin{align}
  & e^{2(\lambda-\lambda_{\rm Schw})}\left[(u^r)^2+r(r-2M)(u^\theta)^2\right]=
    {\cal E}^2-({\cal V}_{\rm eff})^2 \,, \\
  & ({\cal V}_{\rm eff})^2:=
    \left(1-\frac{2M}{r}\right)\!
    \left(1+\frac{\ell^2 e^{2\nu_{\rm disc}}}{r^2\sin^2\theta}\right)
    e^{2\nu_{\rm disc}} \,,
\end{align}
where $\lambda$ has to be computed by line integration of the gradient of total potential $\nu$, with $\lambda_{\rm Schw}$ denoting its pure-Schwarzschild form, $u^\mu$ is four-velocity of the test particle, and ${\cal E}:=-u_t$ and $\ell:=u_\phi$ are constants of the geodesic motion following from the Killing symmetries (they represent specific energy and specific azimuthal angular momentum of a test particle with respect to infinity).
The figure confirms that the pseudo-potentials we consider here provide similar -- but not the same -- effective potentials as the exact Schwarzschild-field description, with the Paczy\'nski--Wiita and our logarithmic formulas apparently being quite close to each other.

Superposition with the Bach--Weyl ring is acquired in the same manner, just with $\nu_{\rm ext}$ represented by
\begin{equation}  \label{BW-ring}
  \nu_{\rm BW} = -\,\frac{2{\cal M}}{\pi\,\sqrt{(\rho+b)^2+z^2}}\;
                  K\!\left(\frac{2\,\sqrt{\rho\,b}}{\sqrt{(\rho+b)^2+z^2}}\right),
\end{equation}
where
$K(k):=\int_0^{\pi/2}\frac{{\rm d}\phi}{\sqrt{1-k^2\sin^2\phi}}\,$
is the complete Legendre elliptic integral of the 1st kind
and ${\cal M}$ and $b$ are mass and Weyl radius of the ring.

\section{Numerics}
\label{numerics}

Trying to check our previous results also by using a different numerical method(s), we turned to symplectic integrators, suitable for conservative systems. However, the two outer sources we consider differ in what to do when the particle hits them: the ring is a curvature singularity, so it is appropriate to halt the trajectory if it gets to its closest vicinity, whereas the thin disc is only singular at its inner edge while cross transitions elsewhere are approximated as non-collisional (pure gravitational effect). Hence, the disc case has to be treated more carefully, regarding that there is a normal-field jump across the equatorial plane (hence jump of the $z$ component of acceleration) above the disc inner radius.

More specifically, the geodesics in the fields given by superpositions with the Bach--Weyl ring are integrated using the 6th-order explicit symplectic partitioned Runge--Kutta method with coefficients adopted from \cite{Yoshida-90} (Solution A) and with step $h=(2\div 5)\cdot 10^{-2}M$ depending on the strength of the ring.

In the case of thin discs (1st inverted Morgan--Morgan disc in our case here), regular integrators bring linear to polynomial growth of error in constants of the motion due to the jump in vertical field. In previous papers of this series, we got over this by the Hut'a method with adaptive step and using higher float representation.
For the present paper, we developed a different variable-step integrator largely inspired by the IGEM code of \cite{SeyrichLG-12} and having the desirable properties of reversibility and symmetry (see \citealt{Stoffer-95} for other varible-step symmetric-reversible integration methods). It is based on Gaussí collocation method with 3 collocation points ($s=3$) and step size determined by the collocation points. Unlike in IGEM, the step size is not determined by the Jacobian of the integrated vector field but by spatial coordinates and the integrated vector field itself (i.e. by phase-space variables and their time-derivatives at the collocation points).

We start by choosing the step
\begin{equation}  \label{origStep}
  h_0 = \frac{\epsilon}{||{\boldsymbol f}_{(1)}+{\boldsymbol f}_{(s)}||} \;,
\end{equation}
where ${\boldsymbol f}_{(i)}:={\boldsymbol f}({\boldsymbol x}_i)$ is the integrated vector field at points ${\boldsymbol x}_i$ of the Gaussian collocation and the norm is defined by $||{\boldsymbol f}||:=\sum |f^j|$, where $f^j$ are components of the vector ${\boldsymbol f}$. (Any norm actually works. We use absolute value which is computationally less demanding than fractional powers, for example). The integrator will be reversible and symmetric if
$h({\boldsymbol x}_i)\equiv h({\boldsymbol p}_1,\dots,{\boldsymbol p}_s;{\boldsymbol q}_1,\dots,{\boldsymbol q}_s)$
is a function symmetric with respect to the reversal of order, $1\leftrightarrow s$,
and to the change of the sign of momenta, ${\boldsymbol p}_i \to -{\boldsymbol p}_i$.
Now, the step $h_0$ is adapted according to\footnote
{We perform the integration in Euclidean $r\sin\theta$, $r\cos\theta$ (not in the Weyl-type coordinates), so we better introduce $\zeta\equiv r\cos\theta$ $(\neq z)$ to avoid confusion.}
\begin{equation}
  h = \frac{h_0}{n(\bar{\zeta})}
    = \frac{\epsilon}{n(\bar{\zeta})\,||{\boldsymbol f}_{(1)}+{\boldsymbol f}_{(s)}||} \;,
\end{equation}
where
\begin{equation}
  n(\bar{\zeta}) := 1 + \frac{\delta_1}{\delta_2+\bar{\zeta}^2} \;,
  \quad
  \bar{\zeta} := \frac{1}{s} \sum_{i=1}^s \zeta_i \,,
\end{equation}
so it remains about $h_0$ for $\bar{\zeta}^2 \gg \delta_1$, while for $\delta_2+\bar{\zeta}^2<\delta_1$ it is contracted; the factor which multiplies $h_0$ is however never less than $\delta_2/\delta_1$. The coefficients $\delta_1$, $\delta_2$ are set so that the particle travels in a controlled manner as close to the equatorial plane as possible.

Then, from some minimal $\zeta$, the particle is reflected with respect to the equatorial plane: when its $|\zeta|$ falls below some chosen $\zeta_{\rm min}$, the program first estimates whether it will cross the equatorial plane in the next $\kappa$ steps by computing
\begin{equation}
  \zeta' = \zeta + \kappa f^\zeta({\boldsymbol x})\,\frac{ \epsilon}{||2{\boldsymbol f}({\boldsymbol x})||} \;,
\end{equation}
thus basically using the Euler explicit method with a step of roughly $\kappa h_0$; if $\zeta$ is found to change sign, the original position is reflected by $\zeta \to -\zeta$. The advantage of this approach is that the particle encounters a ``stepping wall" near $\zeta=0$, the iterative Gaussian collocation does not suffer from the nearby discontinuity and the $\zeta \to -\zeta$ reflection exactly conserves energy. The only point violating the integrator's symmetry is the step estimate of the crossing, but any symmetric reversible stepping would be implicit and difficult to iterate over the discontinuity, with only small benefit to accuracy. We checked that when the parameters are tuned properly, the error typically oscillates without any drift, as typical for symplectic/reversible-symmetric integrators. In some cases the self-adjustment of the step has proven insufficient and a slow linear growth in relative energy error was observed (usually for particles infalling onto a black hole), but this error only rarely exceeded $10^{-11}$.
By numerical experiments, we have found the following parameter ranges to be optimal:
\begin{align*}
  \zeta_{\rm min} &= (1\div 5)\cdot 10^{-4}M, \\
  \delta_1 &= (10^{-5}\!\div 10^{-3})\,M^2, \\
  \delta_2 &= (10^{-8}\!\div 10^{-5})\,M^2, \\
  \epsilon &= (5\div 8)\cdot 10^{-2}M, \\
  \kappa &= 1\div 3 \,.
\end{align*}

Let us add that the Gaussian collocation was found by fixed-point iteration and convergence was confirmed by checking the difference between the current set of collocation points ${\boldsymbol x}_i$ and the previous one ${\boldsymbol x}_{i \rm (p)}$, as represented by
\begin{equation}
  \Delta = \sum_{i=1}^s \sum_{j=1}^{2N}\left|x_i^j-x_{i \rm (p)}^j\right|,
\end{equation}
where $N=2$ is the number of degrees of freedom. The iteration was stopped whenever $\Delta<10^{-13}$. Such a tolerance corresponds to an average error of the order of $10^{-14}$ per collocation component, which is about what can practically be achieved, because spatial position (configuration part of ${\boldsymbol x}$) was often larger than 10, the ``distance" $\Delta$ includes subtraction of close numbers and we used double precision which stores about 15 digits.

The Poincar\'e surfaces of section where created of 3600 equatorial-plane intersections, recording transitions in both directions. Whenever the singularities of the central potential or the ring were closely approached, the integration was stopped and restarted again with a nearby initial condition until a sufficient number of points was collected. However, the whole set of intersections generated by a given trajectory was discarded if a relative error in energy turned out to be too large (namely $\gtrsim 10^{-9}$). Overall, the initial conditions were chosen by a pseudo-random algorithm similar to the one described in paper I.

\begin{figure*}
\includegraphics[width=0.98\textwidth]{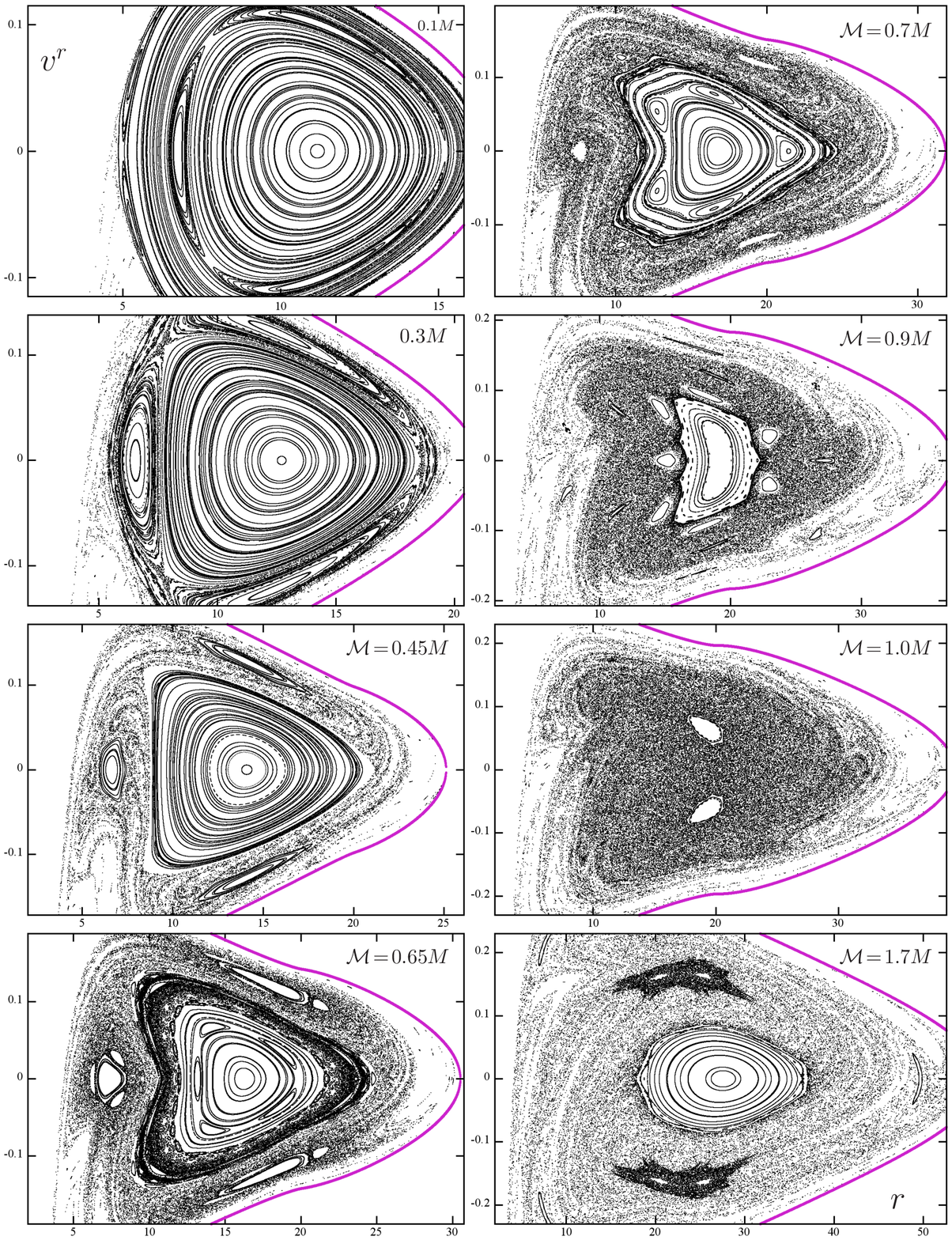}
\caption
{Poincar\'e diagrams in axes $(r,v^r)$ showing passages of geodesic orbits with conserved energy ${\cal E}+1=0.955$ and angular momentum $\ell=3.75M$ through the equatorial plane of a centre described by the Paczy\'nski--Wiita potential (with mass $M$) and surrounded by an iMM1 disc with inner radius $r_{\rm disc}=20M$. Dependence on mass of the disc ${\cal M}$ is shown, as given in the plots. Accessible sector is indicated in purple and $r$ axis is in units of $M$.}
\label{iMM1-mass-PW}
\end{figure*}

\begin{figure*}
\includegraphics[width=0.98\textwidth]{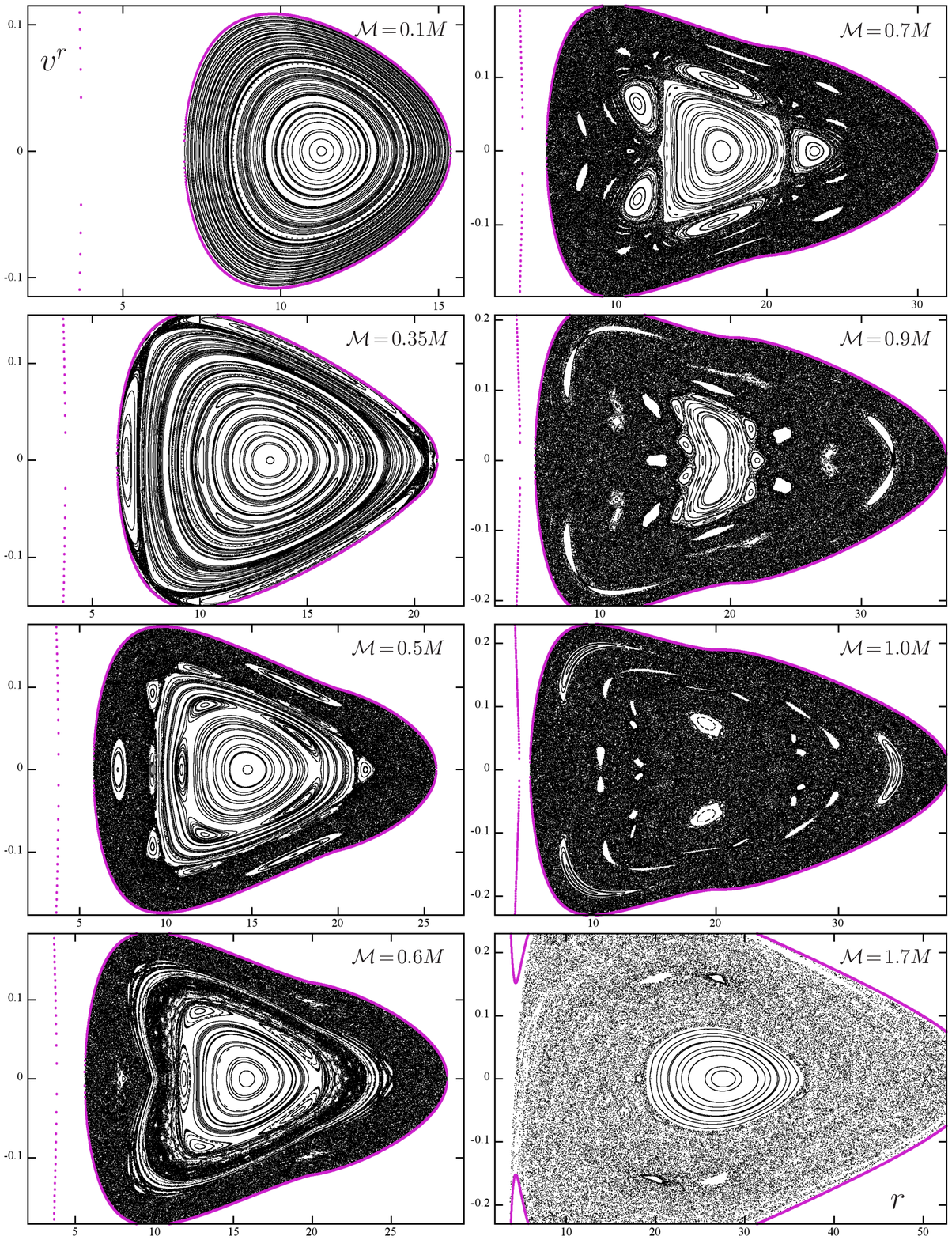}
\caption
{Same series of plots as in Fig. \ref{iMM1-mass-PW}, but with the central black hole simulated by the logarithmic potential (\ref{V_ln}). Comparison of these two figures with fig. 4 of paper I indicates that the phase-space portrait of all the three dynamical systems is similar, though various quantitative differences can be noticed (see mainly behaviour of the accessible region) as more discussed in the main text.}
\label{iMM1-mass-log}
\end{figure*}

\begin{figure*}
\includegraphics[width=0.98\textwidth]{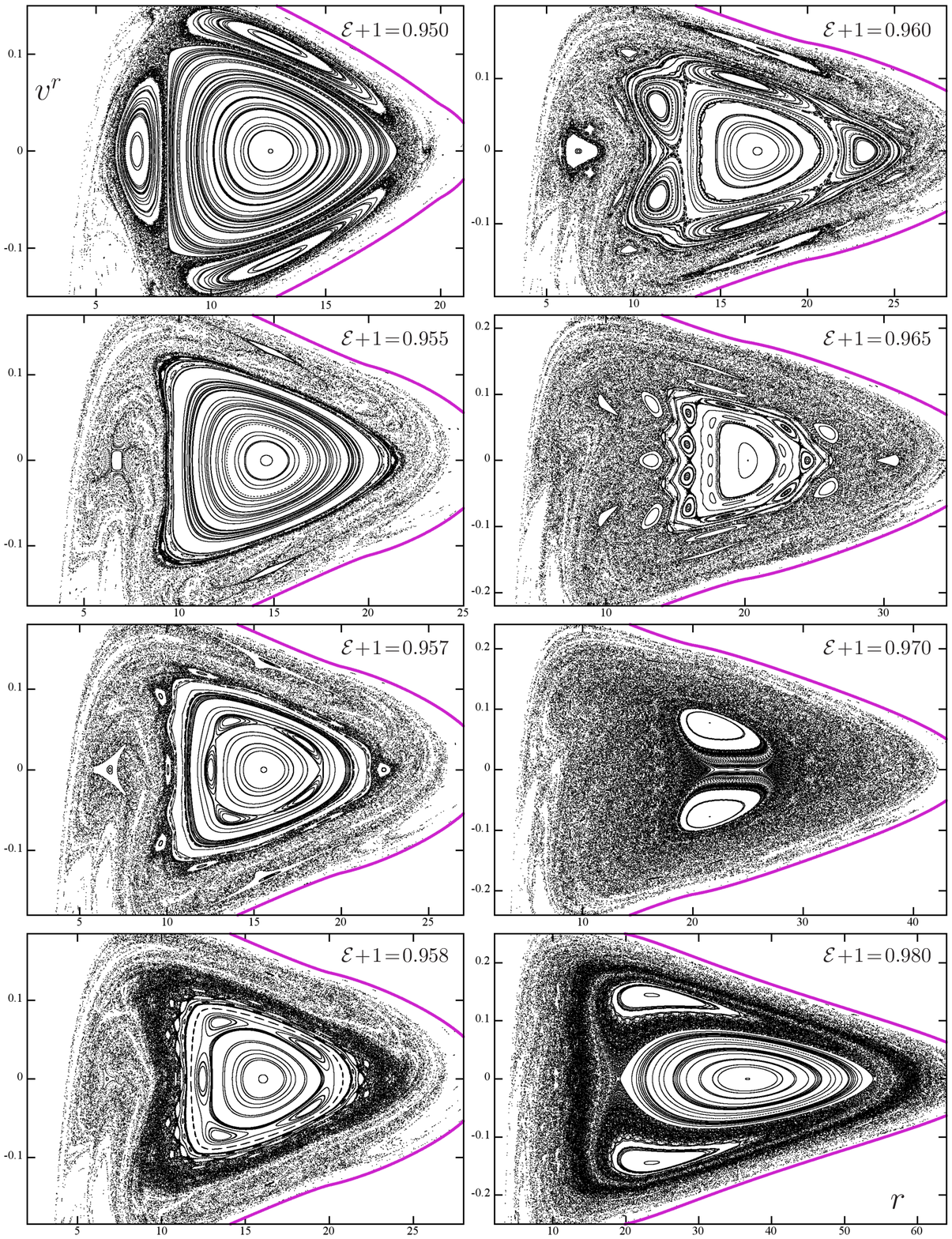}
\caption
{Poincar\'e $(r,v^r)$ diagrams showing passages of geodesics with angular momentum $\ell=3.75M$ through the equatorial plane again, for the centre described by the Paczy\'nski--Wiita potential (with mass $M$) and surrounded by an iMM1 disc with mass ${\cal M}=0.5M$ and inner radius $r_{\rm disc}=20M$. Here dependence on energy of the orbiters ${\cal E}$ is in focus, as indicated in the plots (we enlarge it by unity to match with the relativistic value).}
\label{iMM1-energy-PW}
\end{figure*}

\begin{figure*}
\includegraphics[width=0.98\textwidth]{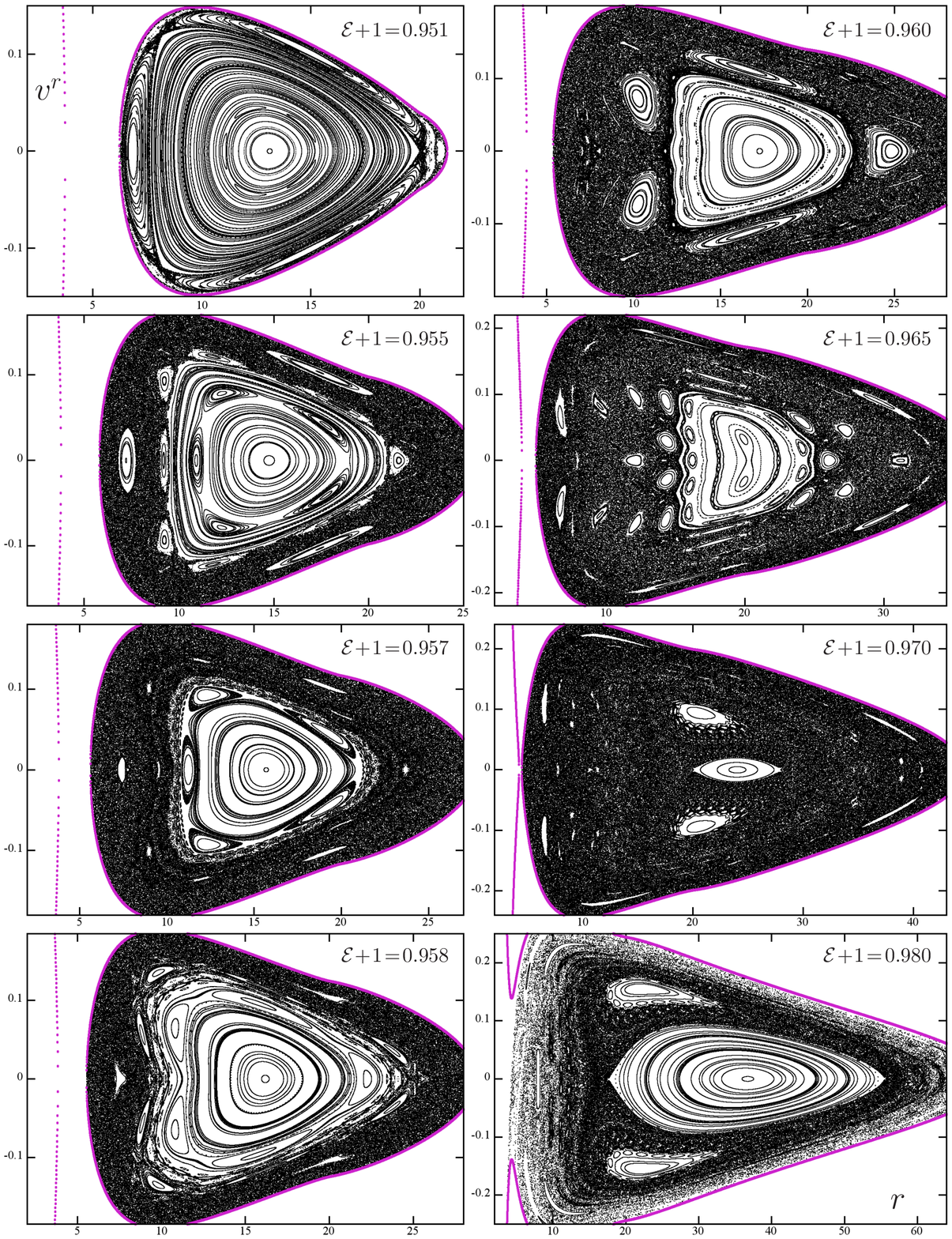}
\caption
{Same series of plots as in Fig. \ref{iMM1-energy-PW}, but with the central black hole simulated by the logarithmic potential (\ref{V_ln}). Comparison of these two figures with fig. 6 of paper I again indicates that both Newtonian dynamical systems well approximate the relativistic one; quantitative differences are further discussed in the main text. (Mainly evident is the different delimitation of accessible phase-space sector again, following from differences in effective-potential profiles.)}
\label{iMM1-energy-log}
\end{figure*}

\begin{figure*}
\includegraphics[width=\textwidth]{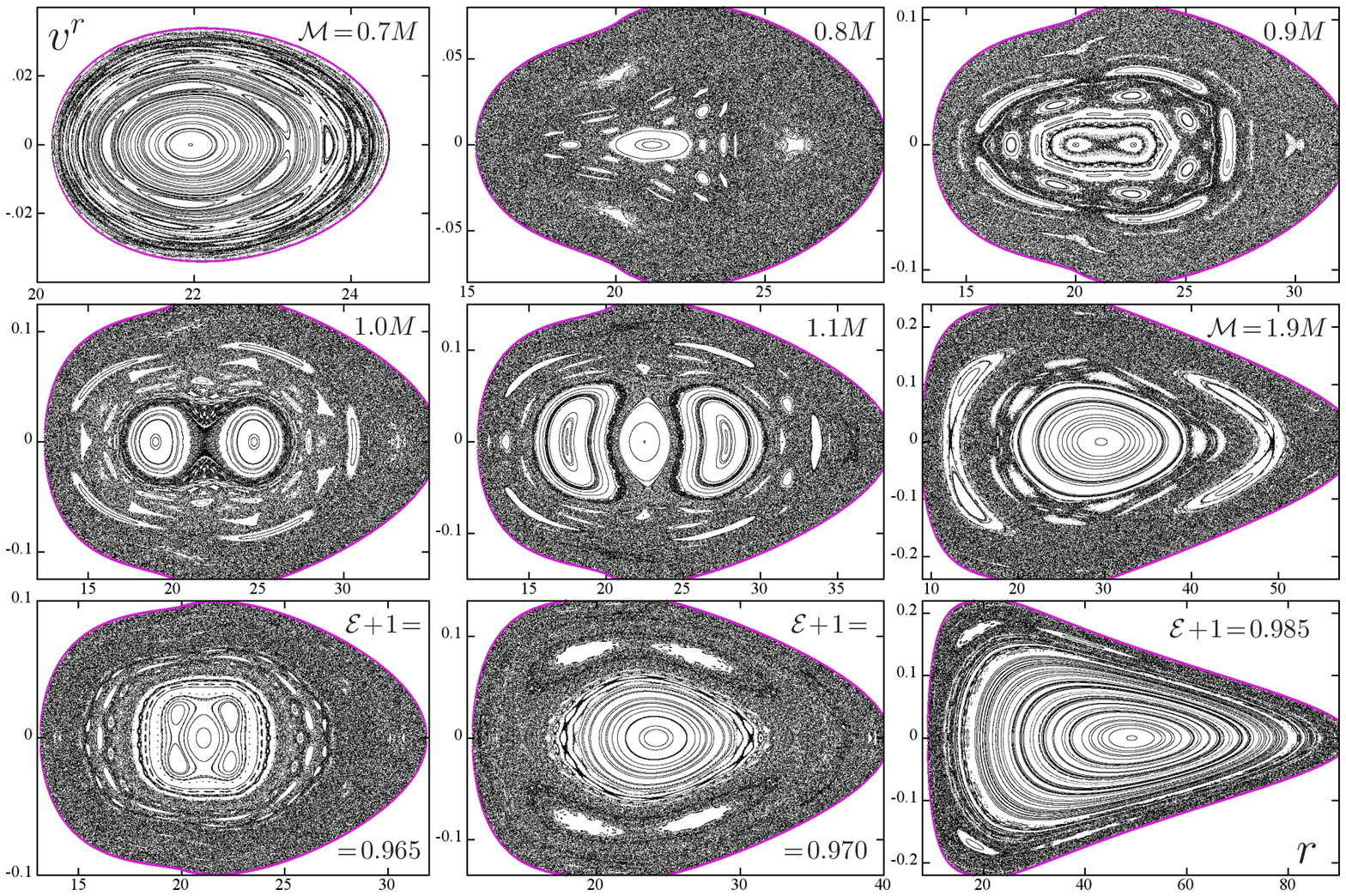}
\caption
{Poincar\'e diagrams in axes $(r,v^r)$ showing passages of geodesic orbits with conserved angular momentum $\ell=3.75M$ through the equatorial plane of a centre described by the Nowak--Wagoner potential (with mass $M$) and surrounded by an iMM1 disc with inner radius $r_{\rm disc}=20M$.
The first two rows show dependence on mass of the disc ${\cal M}$, while all the orbiting particles have energy ${\cal E}+1=0.955$. The last row shows just 3 examples of how the plots change with orbital energy ${\cal E}$, while the disc mass is set at ${\cal M}=0.5M$. The plots are rather different from those involving the Paczy\'nski--Wiita or the logarithmic potential, because the Nowak--Wagoner potential is so weak that it does not form ``its own" valley and the accessible region is maintained by the disc, at least in the ${\cal M}$-series plots. On the other hand, exactly due to this different character it is useful to include at least this one series employing the NW potential.}
\label{iMM1-NW}
\end{figure*}

\begin{figure*}
\includegraphics[width=0.98\textwidth]{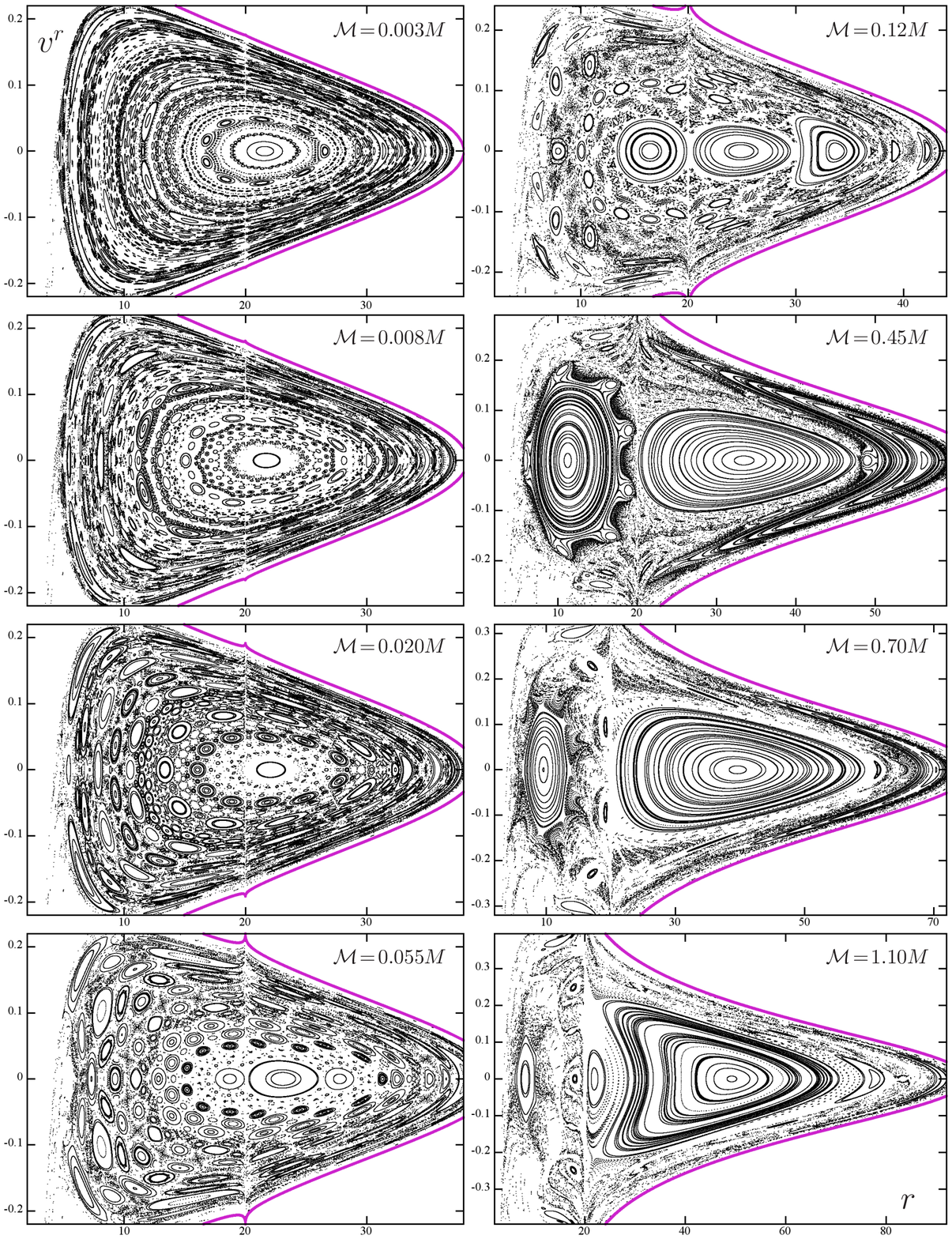}
\caption
{Poincar\'e $(r,v^r)$ diagrams showing passages of geodesics with conserved energy ${\cal E}+1=0.977$ and angular momentum $\ell=3.75M$ through the equatorial plane of a centre described by the Paczy\'nski--Wiita potential (with mass $M$) and surrounded by a Bach--Weyl ring with radius $r_{\rm ring}=20M$. Dependence on mass of the ring ${\cal M}$ is shown, with values given in the plots. Accessible sector is indicated in purple and $r$ axis is in units of $M$.}
\label{BW-mass-PW}
\end{figure*}

\begin{figure*}
\includegraphics[width=0.98\textwidth]{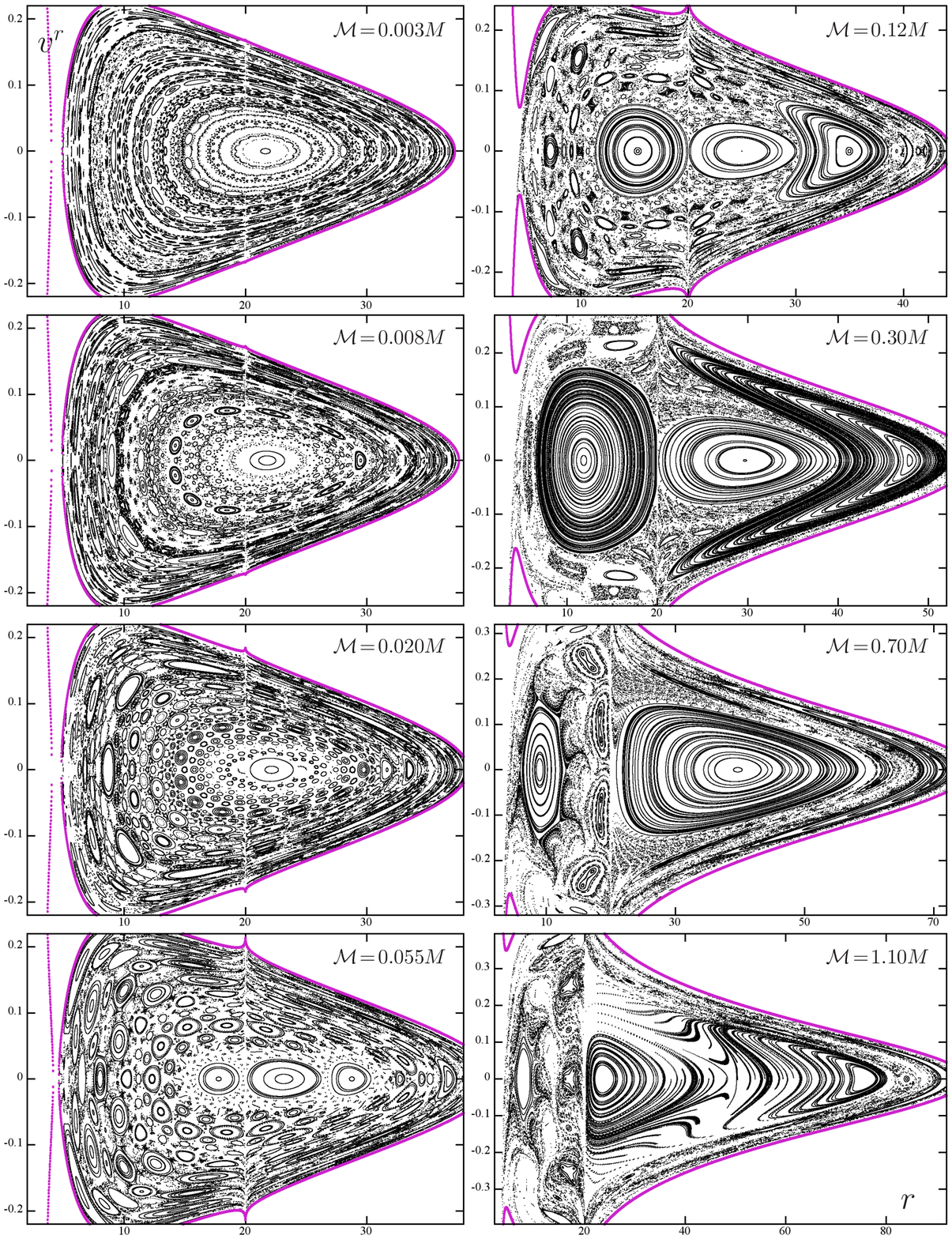}
\caption
{Same series of plots as in Fig. \ref{BW-mass-PW}, but with the central black hole simulated by the logarithmic potential (\ref{V_ln}). Comparison of these two figures with fig. 10 of paper I indicates that the phase-space portrait of all the three dynamical systems is similar, though many quantitative details are different (it would be pointless to discuss them extensively due to the richness of the structure); note again the different delimitation of the accessible region.}
\label{BW-mass-log}
\end{figure*}

\begin{figure*}
\includegraphics[width=\textwidth]{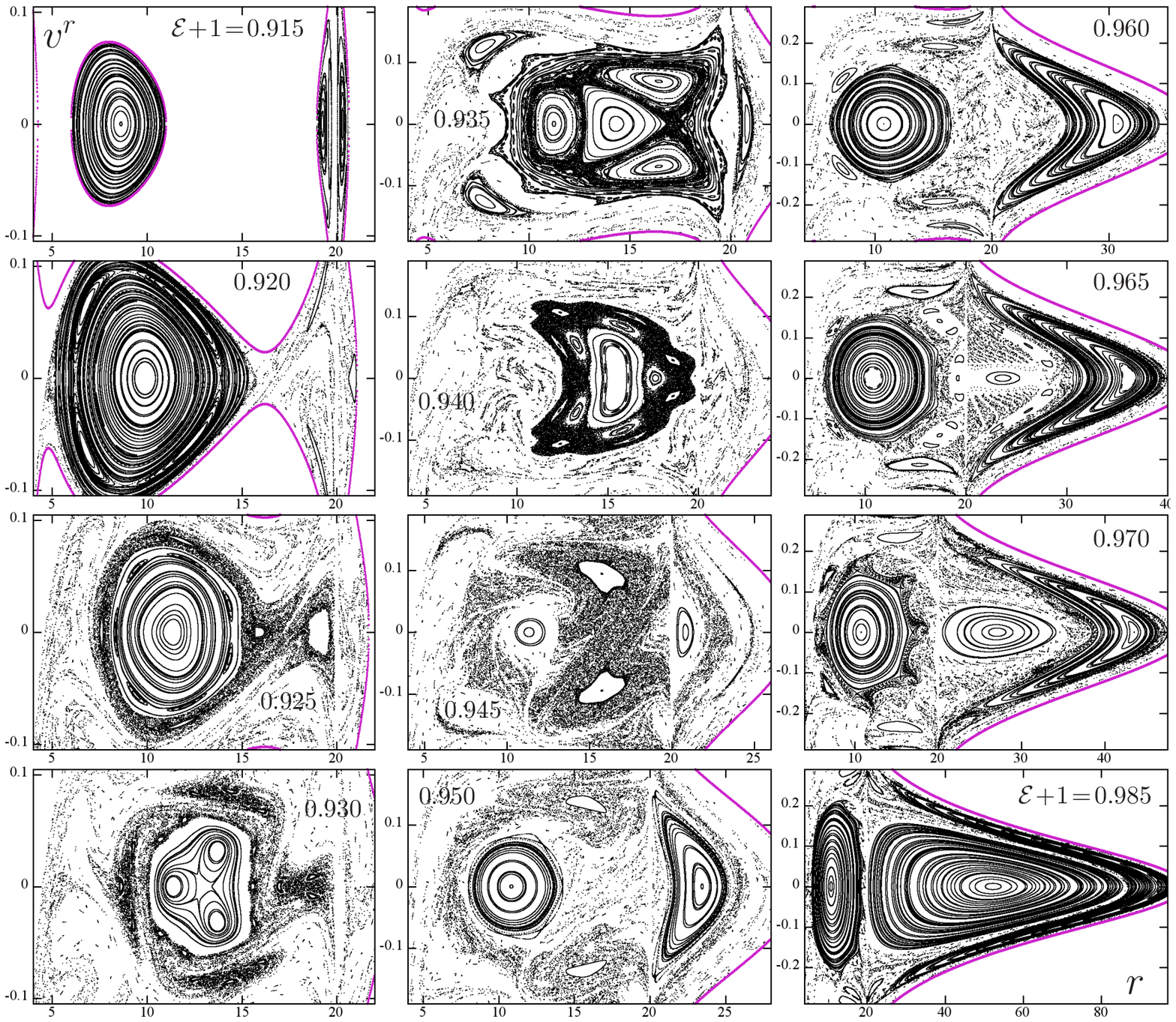}
\caption
{Poincar\'e $(r,v^r)$ diagrams showing passages of geodesics with angular momentum $\ell=3.75M$ through the equatorial plane again, for the centre described by the Paczy\'nski--Wiita potential (with mass $M$) and surrounded by a Bach--Weyl ring with mass ${\cal M}=0.5M$ and radius $r_{\rm ring}=20M$. Here dependence on energy of the orbiters ${\cal E}$ is in focus, as indicated by its values given in the plots (we enlarge it by unity again). We are not showing plots obtained for ${\cal E}+1=0.910$ and less which only contain a tiny accessible region around the ring (the other region between the centre and the ring is not existing yet).}
\label{BW-energy-PW}
\end{figure*}

\begin{figure*}
\includegraphics[width=\textwidth]{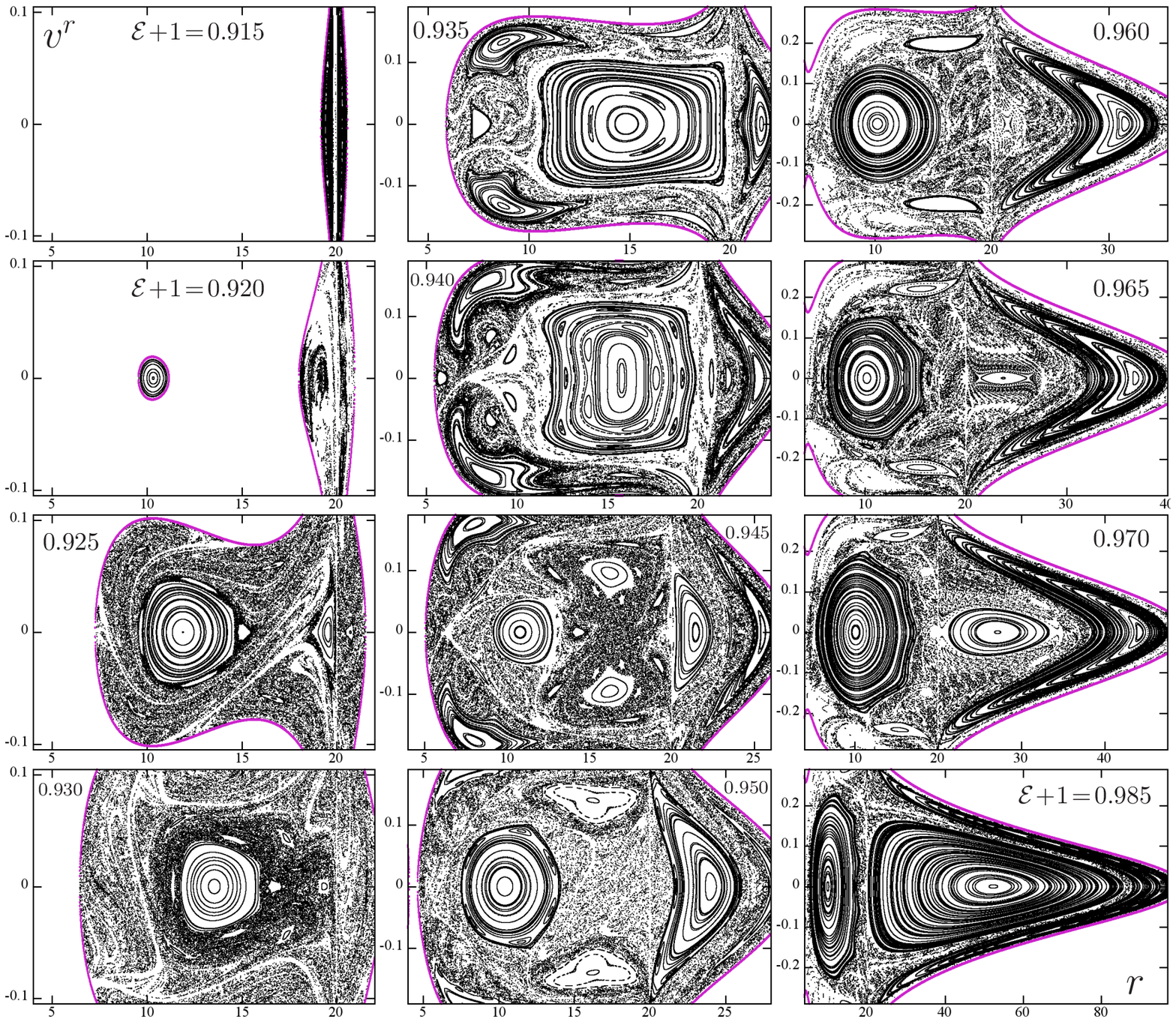}
\caption
{A counterpart of Fig. \ref{BW-energy-PW}, showing the same series of dependence-on-energy plots for the centre described by the logarithmic potential (\ref{V_ln}). All the parameters are kept from previous figure, i.e. $\ell=3.75M$, ${\cal M}=0.5M$, $r_{\rm ring}=20M$, and also the values of ${\cal E}$ are chosen equally, as indicated in the plots. In addition, we have kept exactly the same axis ranges, so the two figures can be compared easily. Their relativistic counterpart is fig. 12 of paper I.}
\label{BW-energy-log}
\end{figure*}

\section{Comparing exact relativistic and pseudo-Newtonian picture}
\label{pseudo-exact-comparison}

Let us stress once more that the (pseudo-)Newtonian and relativistic dynamical systems in question are fundamentally different, because they live in a different configuration space and their evolution is described by a different dynamics as well. It is even impossible to decide which situations are ``similar", because most of the relevant variables actually have different meaning within Newtonian and relativistic case; for example, if one places the external source at some ``given" radius, it has different meanings in the Euclidean spherical/cylindrical coordinates $r\sin\theta$, $r\cos\theta$, in the Weyl-type coordinates $\rho=\sqrt{r(r-2M)}\,\sin\theta$, $z=(r-M)\cos\theta$ (which we use here) and -- in relativity -- in terms of proper radial distance or in circumferential radius. Therefore, one can only wish for a reasonable correspondence of {\em qualitative} phase-space features and of their evolution with analogous parameters. Yet it will still be interesting to see whether and which of the potentials reproduce at least some of the quantitative aspects, like the pattern of resonances and the sequence of their appearance.

Needless to say, one has only a very restricted space here for such a comparison. It is symptomatic for non-integrable systems that their dependence on parameters is ``chaotic" (non-smooth) -- they may change only slowly within one parameter range, whereas very abruptly within the other (which may be quite narrow). Being only able to select several {\em sections} through the very rich parameter space of the systems, one can either take those with the same values of the corresponding parameters, those showing similar features, or simply those where something interesting is happening. Without adhering to any strict rule, we have generally set the fixed parameters at formally identical values as in the relativistic case and varied the free parameter in roughly the same range.

The comparison in general reveals that the overall tendency is the same in both the relativistic and pseudo-Newtonian systems: when the perturbation strength (disc mass in our case) or particle energy increases, the system first gets more and more chaotic, whereas for very large parameter values the ``primary" regular region slowly grows again. However, since such a behaviour is quite typical for weakly non-integrable systems, we will mainly try to note the differences.

We start by evolution of the phase portrait with mass of the external source.
Figures \ref{iMM1-mass-PW} and \ref{iMM1-mass-log} show how Poincar\'e diagram of equatorial transitions changes with relative mass of the inverted 1st Morgan-Morgan disc as the external source. Placing the inner edge of the disc on $r_{\rm disc}=20M$ and setting ${\cal E}=0.045\,(=0.955-1.000)$, $\ell=3.75M$ as in paper I, the figures present diagrams obtained for 8 different values of ${\cal M}/M$ between $0.1M$ and $1.7M$. The Schwarzschild centre is imitated by the Paczy\'nski--Wiita potential in Fig. \ref{iMM1-mass-PW} while by the logarithmic potential in Fig. \ref{iMM1-mass-log}. We have not included the Nowak--Wagoner potential in the detailed study, because it has turned out to yield rather different results, not well compatible with the exact relativistic picture (the NW potential is ``too weak" and for a large portion of the studied parameter ranges its phase space bears no bound particles); however, Fig. \ref{iMM1-NW} is provided for cursory illustration.

\begin{figure}
\includegraphics[width=\columnwidth]{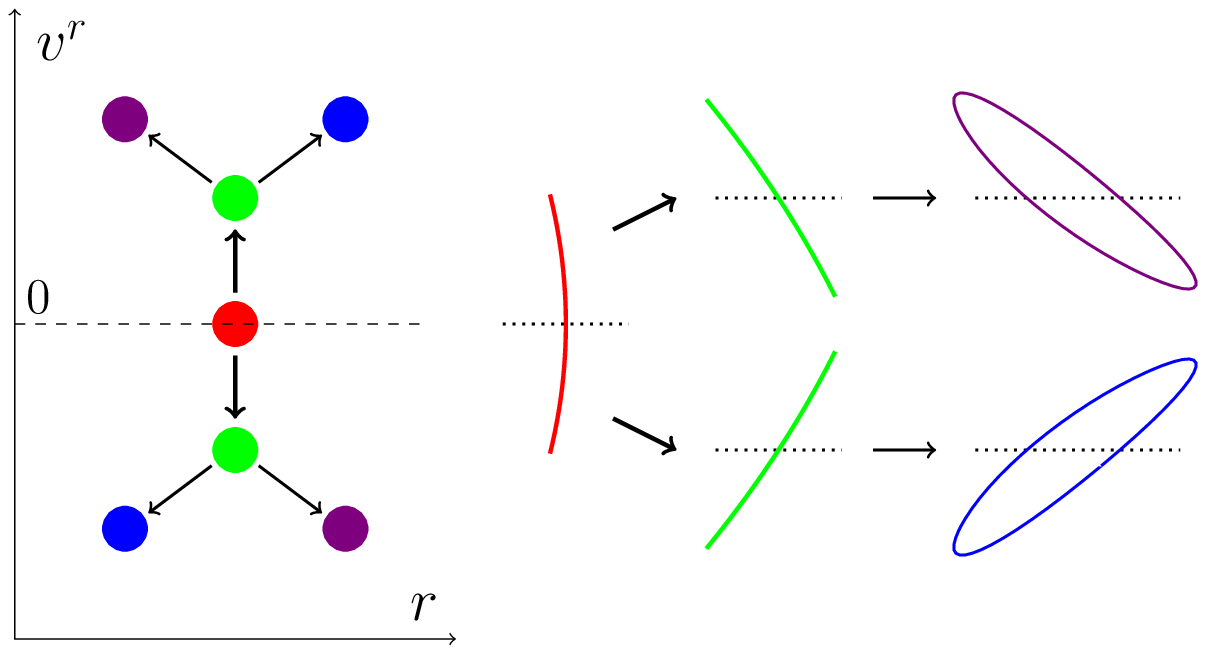}
\caption
{A scheme of the 1:1-resonance bifurcations occurring both in the relativistic and pseudo-Newtonian hole-disc field. The left part indicates the imprints of the trajectories on the Poincar\'e section, while the right part illustrates the corresponding trajectory shapes (coloured respectively) within the $(r\sin\theta,r\cos\theta)$ plane, with the dotted line always representing the equatorial plane $\theta=\pi/2$. Note that the sections of the blue and purple trajectories would in fact depend on the direction of velocity on the curve. The original 1:1 island first breaks up ``vertically" and then both new islands further decouple ``horizontally". This typically happens in stage when the phase space is the most chaotic, which in terms of the disc mass as the perturbing agent means ${\cal M}\sim M$. In the present paper, the three phases can be seen in Figs. \ref{iMM1-mass-PW} and \ref{iMM1-mass-log}, in plots with ${\cal M}=0.9M$, $1.0M$ and $1.7M$; the corresponding relativistic situations were plotted in fig. 4 of paper I (see the plots with ${\cal M}=1.0M$ and $1.1M$ there; the 3rd phase was not shown).}
\label{bifurcation}
\end{figure}

The Figs. \ref{iMM1-mass-PW} and \ref{iMM1-mass-log} are to be compared with fig. 4 of paper I.
\begin{itemize}
\item
The main difference concerns the accessible domain which, in comparison with the exact relativistic case, is more open towards the centre for the Paczy\'nski--Wiita potential, whereas more closed for the logarithmic potential (see Fig. \ref{ell-on-En}): for real Schwarzschild, the domain is closed first, enlarges with increasing disc mass and finally opens towards the centre when the disc mass reaches about half of the black-hole mass (this applies specifically to the iMM1 disc with $r_{\rm disc}=20M$, of course). In contrast, for the Paczy\'nski--Wiita potential the accessible sector is always open towards the centre, whereas for the logarithmic potential it is closed and only opens after the disc outweighs the centre. However, this does not seem to be that crucial for evolution of the phase-space features, the opening only enables the centre to ``suck out" the outer chaotic sea. (This makes the open diagrams asymmetric with respect to $v^r=0$.)
\item
The similarity of all three systems is really striking, with most phase-space structures appearing and in the same succession. In the Paczy\'nski--Wiita case, similar features appear for somewhat lower disc-mass values (about $0.1M$$\div$$0.2M$ ``earlier") than in the relativistic case, while for the logarithmic potential they appear still about $0.05M$$\div$$0.1M$ earlier than for the PW potential. This may be interpreted as slightly stronger inclination of the pseudo-Newtonian system towards chaos, which is in accord with our preliminary guess stemming from deeper potential valleys provided by them (see Fig. \ref{eff-potentials-1valley}).
\item
More details about the structures: with increasing perturbing mass, the relativistic geodesic system (see paper I) first develops a 3-fold island within the primary regular region (2:3 resonance, ``fish"-shaped orbit in Fig. \ref{orbit-shapes}); then (temporarily) a 4-fold one appears within the chaotic periphery of the accessible region: this is a particularly shaped ``symmetrized set" of 1:2 resonances (analogous feature appears ``earlier" in the PW case).\footnote
{Normally, an $m$:$k$ resonance is associated with a $k$-fold ($k$-periodic) island. It is not clear whether the 4-fold island represents a tangent or a pitchfork bifurcation of the 1:2 resonance (cf. also the following commentary on a 1:1-resonance bifurcation).}
Later the central regular region gives birth to five islands (4:5 resonance, again identical in the relativistic and pseudo-Newtonian case), then even 7-fold and 9-fold ``baby-islands" (6:7 and 8:9 resonances) can be spotted, and finally the region breaks up into two parts symmetrical with respect to $v^r=0$ which disappear shortly after the disc mass reaches about the black-hole mass. Meanwhile, a central regular sector appears and grows gradually with the disc mass increased still more.\\
The Paczy\'nski--Wiita centre with the iMM1 disc also first give birth to the 3-fold island and then to the 5-fold, 7-fold and 9-fold ones, corresponding to the same resonances as in the relativistic case; the 4-fold structure only appears in a light-disc stage (along the border of the regular domain).
The logarithmic potential yields very similar behaviour, with the 4-fold structure not occurring at all.
\item
The breakup of the original central island is a very characteristic feature of the relativistic as well as of the pseudo-Newtonian systems; in all cases it occurs when the disc mass ${\cal M}$ reaches about that of the central hole ($M$). More specifically (Fig. \ref{bifurcation}): if one takes any point $r$, $\theta$, $\dot{r}$, $\dot{\theta}$ on the original central orbit (red) and applies the reflection $\theta\to\pi-\theta$ and/or velocity-reversal $\dot{r}\to -\dot{r}$, $\dot{\theta}\to -\dot{\theta}$, the same central orbit is obtained, just in a different phase. Namely, the central orbit is -- up to a phase shift -- symmetric with respect to reflection and reversal which are discrete symmetries of the Hamiltonian. However, this symmetry of the whole phase space need not be respected by {\em individual} invariant structures. The multiplication of resonant islands is then a kind of ``spontaneous symmetry breaking", because as the central orbit shifts to the strongly perturbing disc edge, it looses stability and bifurcates into two (green) orbits which are reflection-symmetric when taken together as a ``symmetrized set" (the reflection operation maps the points of the first trajectory on the second one and vice versa). These green trajectories later bifurcate even further in the radial direction, into 2+2 ``reversible-asymmetric" trajectories (blue and purple). The four small islands in the Poincar\'e diagrams with ${\cal M}=1.7M$ in Figs. \ref{iMM1-mass-PW} and \ref{iMM1-mass-log} thus correspond to a symmetrized set of four distinct 1:1 resonances.
\item
Let us point to one specific difference finally:
In the log-potential system, one observes a strong 5-fold structure corresponding to a 4:5 resonance inside the central regular region, existing from ${\cal M}=0.33M$ to ${\cal M}=0.62M$ (we mean the one oriented so that one ``vertex" island lies on the $v^r=0$ axis and  towards the centre); in the PW-potential system, the similar structure is weaker and only persists from ${\cal M}=0.54M$ to ${\cal M}=0.67M$; in the exact system it does not appear at all. (However, it {\em can} appear rarely for a different type of disc and/or for a disc placed on different radius -- see fig. 5 in paper I, plots with $r_{\rm disc}=14M$ and $15M$.) Notice also how in the pseudo-Newtonian cases that structure finally switches over to a complementary/reverse 5-fold pattern, with one ``vertex" lying {\em away} from the centre (which {\em is} common in the exact system).
\end{itemize}

Now we proceed to energy which is one of the most important parameters of any dynamical system.
Figures \ref{iMM1-energy-PW} and \ref{iMM1-energy-log} show how Poincar\'e diagram of equatorial transitions changes with conserved energy of the freely orbiting test particles. Placing the ``iMM1" disc of mass ${\cal M}=0.5M$ from $r_{\rm disc}=20M$ and setting $\ell=3.75M$ as in paper I again, the figures present diagrams obtained for 8 different values of ${\cal E}$ between ${\cal E}+1=0.95$ and ${\cal E}+1=0.98$. The Paczy\'nski--Wiita potential is used in Fig. \ref{iMM1-energy-PW} while the logarithmic potential in Fig. \ref{iMM1-energy-log}. The Nowak--Wagoner potential is only illustrated briefly in Fig. \ref{iMM1-NW}.

The Figs. \ref{iMM1-energy-PW} and \ref{iMM1-energy-log} are to be compared with fig. 6 of paper I. The latter shows less stages than we present here, but the comparison anyway confirms what has already been observed above in figures illustrating dependence on the perturbing mass (the sequences in fact resemble the previous ones): the pseudo-Newtonian systems well simulate the exact relativistic one, they are just slightly richer of tiny structures and display major features somewhat ``earlier" in terms of the relevant parameter (here energy). In this sense, they can again be called ``more chaotic" than the exact system, with the logarithmic potential perhaps being slightly more prone to irregularity than the Paczy\'nski--Wiita one. In the left column of Fig. \ref{iMM1-energy-log}, notice the nice (center-vertexed) 5-fold pattern and its switch-over to the ``complementary" pattern between ${\cal E}+1=0.957$ and ${\cal E}+1=0.958$.

The same kind of illustration -- dependence on external mass and on orbital energy -- is also provided for the black-hole--like centre surrounded by a Bach--Weyl ring (with radius $r_{\rm ring}=20M$). Figures (\ref{BW-mass-PW}) and (\ref{BW-mass-log}) show how the equatorial $(r,v^r)$ section through the phase-space evolves with relative mass of the ring, while energy and angular momentum integrals are chosen ${\cal E}+1=0.977$ and $\ell=3.75M$; the centre is described by the PW potential in Fig. \ref{BW-mass-PW}, while by the ln potential in Fig. \ref{BW-mass-log}. Figures \ref{BW-energy-PW} and \ref{BW-energy-log} show dependence on energy of the orbiting particles, while $\ell=3.75M$ and the ring mass is set at ${\cal M}=0.5M$; again the PW potential is employed in the first figure, while the ln potential is employed in the second one. Figures (\ref{BW-mass-PW}) and (\ref{BW-mass-log}) are counterparts of fig. 10 in paper I, while Figs. \ref{BW-energy-PW} and \ref{BW-energy-log} are counterparts of fig. 12 in paper I.

The comparison with paper I again verifies quite close similarity of all the three dynamical systems (the exact relativistic one and those with the black-hole simulated by the PW or the ln potential). One might however notice many unlike details, but they are not worth careful discussion, because in the ring case the dynamics is apparently very rich of tiny structures, both regular and chaotic. The rich ornamentation follows from close encounters with the singular source, so in future work -- whether within exact description or using pseudo-potentials -- it will be sensible to rather consider orbits {\em not closely interacting with the ring}, i.e. to choose the accessible region so that not to involve the ring radius. Such a configuration will also be more realistic since there are no literally singular sources in nature (cf. paper III where this point was checked in simple modelling of Galactic circumnuclear rings).

Anyway, comparison of Figs. \ref{BW-energy-PW} and \ref{BW-energy-log} with fig. 12 of paper I indicates, similarly as the centre-disc plots above, that the pseudo-Newtonian imitations of black hole lead to slightly faster ``evolution" with parameters than the exact relativistic case, which can perhaps be interpreted as more ``unstable" response to the perturbation. For instance, the breakup of the principal regular sector existing below the ring occurs at ${\cal E}=0.945\div 0.950$ in the relativistic case, while at ${\cal E}+1=0.940\div 0.945$ in both pseudo-Newtonian cases.
Again quite different is the moment of opening of the accessible domain towards the centre: in the relativistic picture, this happens at ${\cal E}\simeq 0.934$, while with the PW potential it happens at ${\cal E}+1\simeq 0.917$ (the PW potential is almost ever open) and with the ln potential it happens only at ${\cal E}+1\simeq 0.953$ (the ln potential is almost ever closed).

\begin{figure}
\includegraphics[width=\columnwidth]{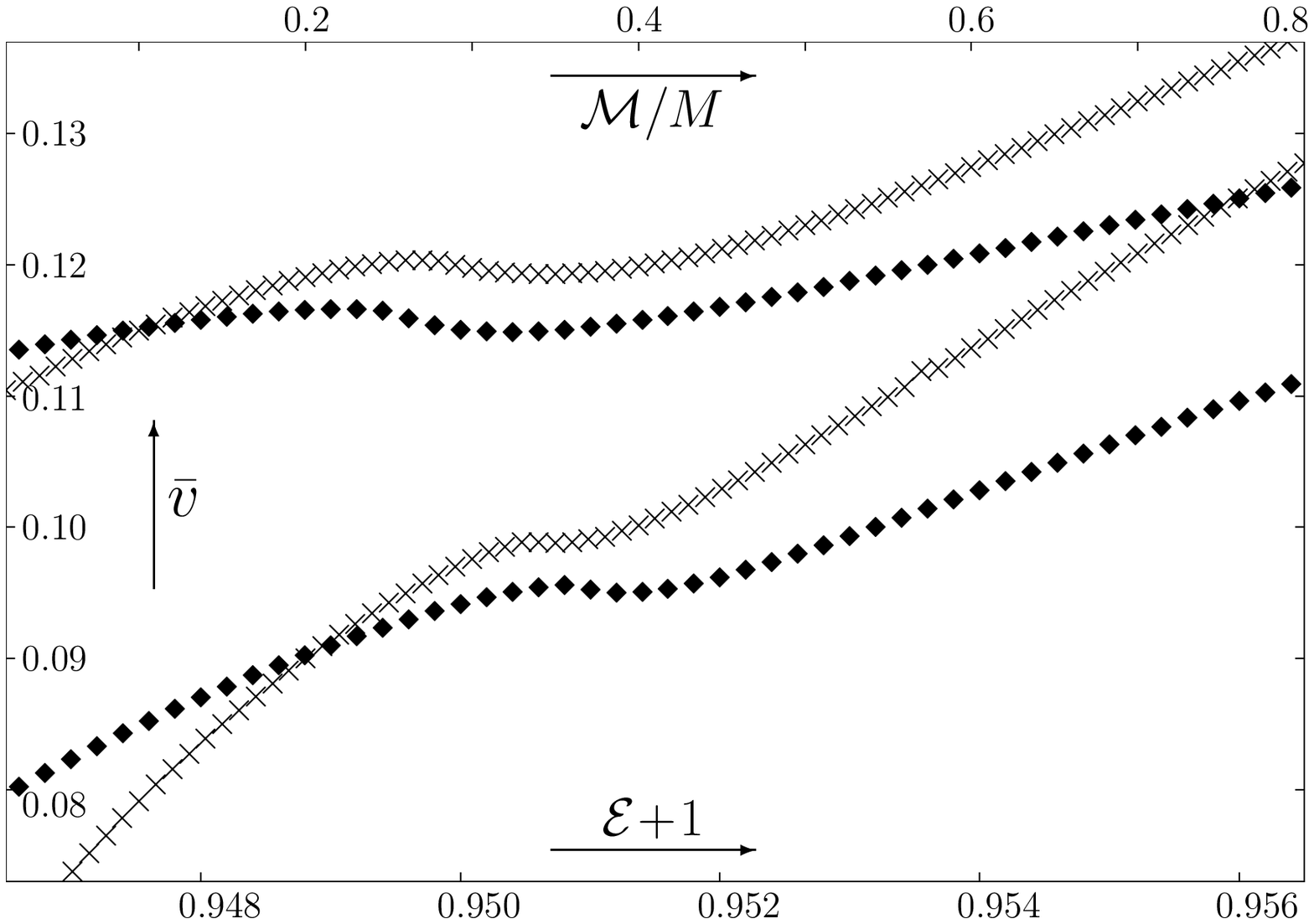}
\caption
{Average speed (\ref{bar{v}}) with which the orbits (having $\ell=3.75M$) intersect the equatorial plane of the system of a black hole surrounded by the iMM1 disc with inner radius $r_{\rm disc}=20M$. The dependence of $\bar{v}$ on the relative disc mass ${\cal M}/M$ is plotted for ${\cal E}+1=0.955$ (these curves are given in $\times$ crosses; top axis applies); the dependence of $\bar{v}$ on the conserved energy ${\cal E}$ is plotted for the disc mass ${\cal M}=0.5M$ (these curves are given in solid diamonds; bottom axis applies). The top couple of curves has been obtained for the Paczy\'nski--Wiita potential, while the bottom (faster growing) couple for the logarithmic potential. In both cases, $\bar{v}$ grows almost monotonously with ${\cal M}$ as well as with ${\cal E}$, having a single ``dip" which is associated with the phase when the accessible region reaches above the disc edge.}
\label{average-speed}
\end{figure}

\subsection{On dependence on perturbing mass and on orbital energy}

The Newtonian dynamics allows for a straightforward and quantitative explanation of the correspondence between the changes in sections caused by variation of the perturbing mass ${\cal M}$ and by variation of the orbital energy ${\cal E}$. First, as we fix the total energy ${\cal E}$ and increase the disc mass ${\cal M}$, the potential well becomes deeper, so the particle necessarily gets more {\em kinetic} energy. Hence, although the parameters of the surfaces of section in Figs. \ref{iMM1-mass-PW} and \ref{iMM1-mass-log} might look like we study ``identical" ensembles of trajectories subjected to a stronger and stronger dynamical perturbation, effectively it is not so.

To illustrate this point further, we compute the average speed $\bar{v}({\cal E},{\cal M},\ell,r_{\rm disc})$ over the equatorial plane for the parameters chosen in Figs. \ref{iMM1-mass-PW}--\ref{iMM1-energy-log},\footnote
{The ``average speed" is certainly an ambiguous concept. We choose here a definition which is simple and natural.}
\begin{equation}  \label{bar{v}}
  \bar{v}({\cal E},{\cal M},\ell,r_{\rm disc})
  = \frac{\int\sqrt{2\left[{\cal E}-{\cal V}_{\rm eff}(\theta\!=\!\pi/2)\right]}\;2\pi r\,{\rm d}r}
         {\int 2\pi r\,{\rm d}r} \;,
\end{equation}
and plot the dependence of the result on ${\cal M}$ and ${\cal E}$ for the PW and ln potential in Fig. \ref{average-speed}. (Integration is performed over the accessible region; in cases where the the latter was not closed in the direction toward the centre, we have taken the lowest reachable $r$ to be $5M$.) In the ranges $0\lesssim{\cal M}\lesssim 0.8M$ and $0.945\lesssim{\cal E}\!+1\!\lesssim 0.965$ (of which part is shown in the figure), and for both central potentials, the growth of $\bar{v}$ with either ${\cal M}$ or ${\cal E}$ is very similar.
The comparison of plots shown in Fig. \ref{average-speed} thus suggests the following interpretation: the phase-space structure stays roughly the same for a moderate disc-mass perturbation, with the growing disc mass mostly inducing a shift of the orbits to higher kinetic energies. This aspect is surely present in the relativistic case studied in papers I--III as well, but would require a more subtle argument.

\subsection{Remark on the Bach--Weyl ring}

The Bach--Weyl ring is actually an interesting source. Its potential (\ref{BW-ring}) is everywhere attractive, namely its field intensity (minus gradient of the potential) points toward the ring from all local latitudinal directions. In the Newtonian picture it thus represents an ``ordinary" ring source. In relativity the potential remains valid, but the metric involves {\em two} functions, the second being given by a line integral of the potential gradient. In the BW-ring case, both functions are given by elliptic integrals and, as expected, both diverge at the very ring. The two divergences however combine to such a deformation of geometry in the ring's vicinity that the real physical distance (proper distance) to the ring comes out finite from outside (when the ring is approached from bigger radii), whereas infinite from inside (when the ring is approached from smaller radii). When free motion is plotted in coordinates, the particles thus appear repelled/attracted by the ring in the directions from which the ring is physically nearby/far away, i.e., they seem to be repelled towards larger radii, whereas attracted from smaller radii. The effect is strongest in the equatorial region. We noticed it and interpreted in \cite{SemerakZZ-99b}, and later this was repeated by \cite{DAfonsecaLO-05}.

Since the above feature is ``felt" up to several tenths of ring mass in the Weyl or Schwarzschild coordinates (in geometrized units), it might be somehow reflected in orbital statistics. However, the effect is much better seen in the {\em meridional} plane (than in the equatorial one): the coordinate tracks of free particles, when approaching the ring from any latitudinal direction, are driven towards its inner side and hit it just along the equatorial plane. Inspection of the ring's neighbourhood in Poincar\'e plots does not seem to indicate stronger anisotropy in the relativistic case. One can only observe slight differences in evolution of the main regular region centered just above the ring: for small ring mass, it is central symmetric in all three descriptions, but when the mass reaches several percent of $M$, it ``elongates" along the $v^r=0$ direction and finally two new islands establish on its opposite radial sides, created by orbits circling around (``through") the ring. This process starts somewhat before ${\cal M}=0.02M$ in the relativistic system as well as in the system using the ln potential, while in the PW-potential case it starts only before ${\cal M}=0.03M$. The only qualitative difference between the relativistic and the pseudo-Newtonian systems is that in the latter case, for large ring masses (from ${\cal M}=0.8M$ for the ln potential, while from ${\cal M}=0.9M$ for the PW potential) a new pair of regular regions appear, again symmetrically with respect to the principal island, but now both lie {\em above} the ring radius. See mainly the last plot (${\cal M}=1.1M$) of the ln-potential Fig. \ref{BW-mass-log}, where these two islands already dominate the section. It would be interesting to check whether the lack of this regular couple in the relativistic system has connection with the ring's outward repulsion.

However, it should be noted that the Poincar\'e-surface analysis is best suited for the demonstration of long-term effects in the motion of eternally orbiting particles, whereas the above mentioned feature mainly affects trajectories soon to be captured by the ring. Thus, the Poincar\'e section will typically bear one or two points from such trajectories and their dynamical behaviour will be hardly discernible for most part. The only effect one could hope to observe in the surfaces of section is a deformation of invariant structures -- of which we find no persuasive evidence.

\begin{figure}
\includegraphics[width=\columnwidth]{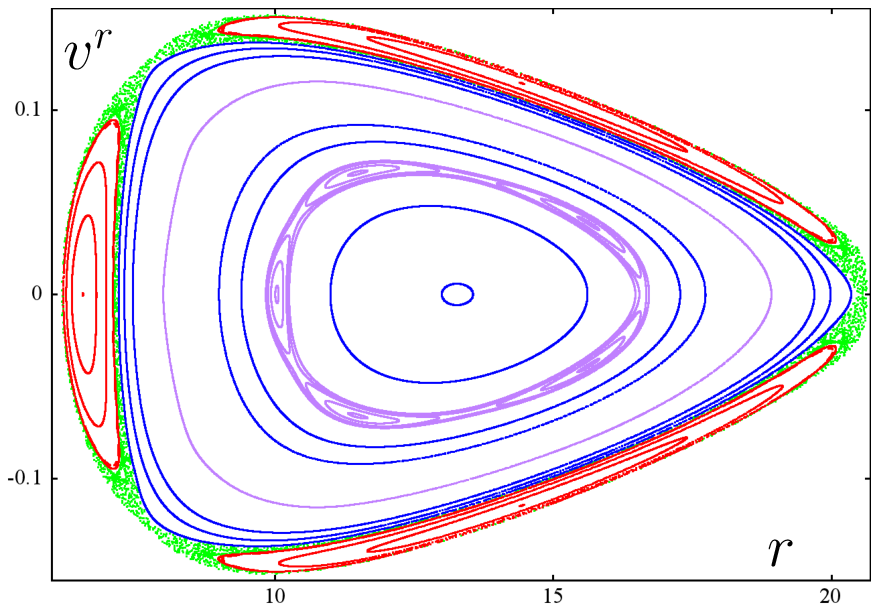}
\caption
{The Poincar\'e diagram ${\cal M}=0.35M$ from Fig. \ref{iMM1-mass-log} revisited with the aim to illustrate what kind of orbits its main structures represent. About 7400 transitions for each orbit has been recorded. The orbits are shown in colour to ensure their easy identification against Fig. \ref{orbit-shapes} where the meridional-plane shapes of their 200 periods are plotted.}
\label{orbit-shapes-Poincare}
\end{figure}

\begin{figure*}
\includegraphics[width=\textwidth]{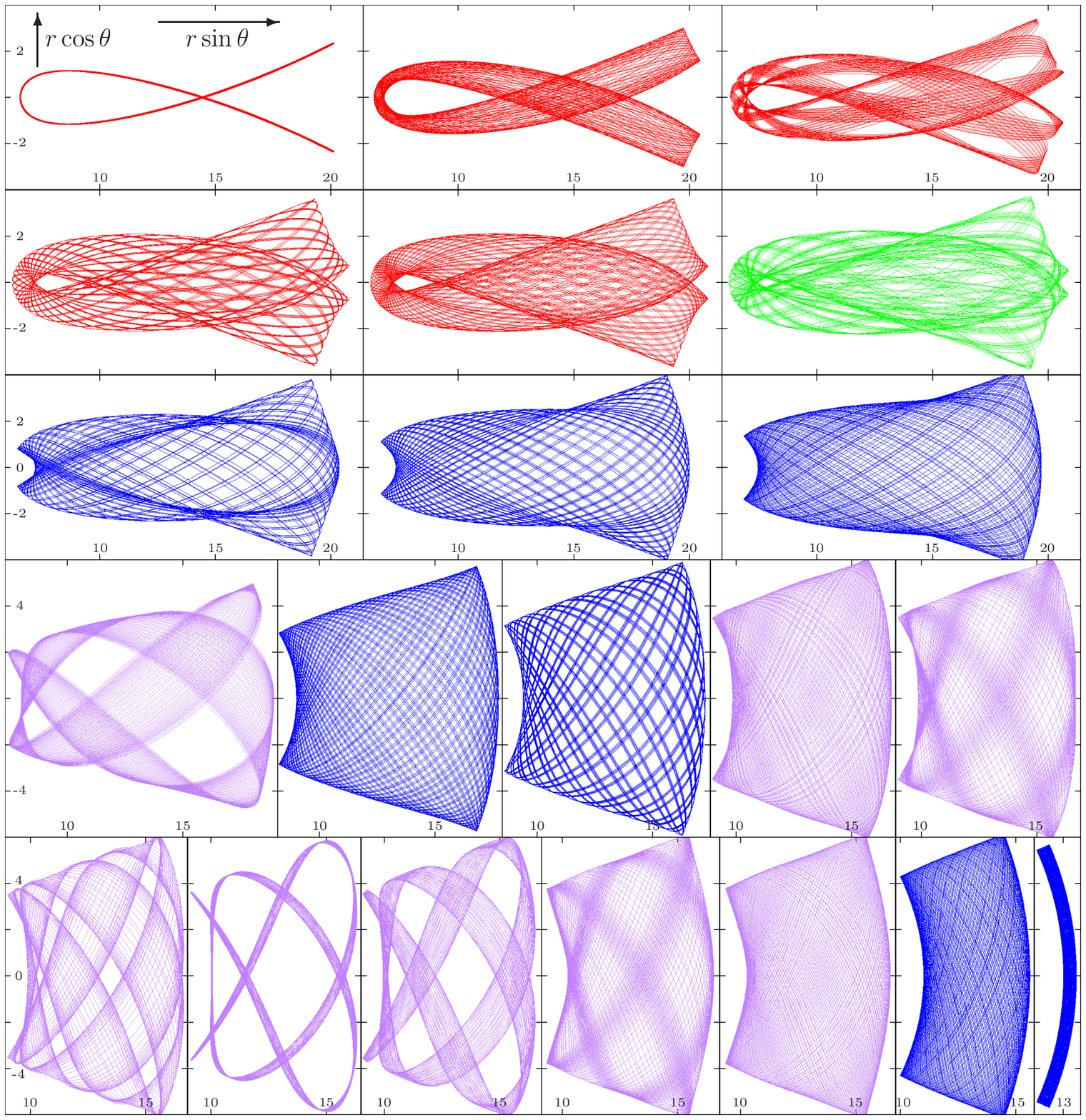}
\caption
{A counterpart of Fig. \ref{orbit-shapes-Poincare}, showing the shapes within the $r\sin\theta$, $r\cos\theta$ plane of the trajectories whose equatorial transitions have been recorded there (in the $r$, $v^r$ axes). The orbits are plotted up to some 200 periods and are coloured to be easily identifiable in Fig. \ref{orbit-shapes-Poincare}. From top left to right and bottom, the profile starts from the central orbit of the 3-fold island and proceeds toward the centre of the Fig.-\ref{orbit-shapes-Poincare} surface of section. All the plots have exactly the same scale, though the coordinate ranges (indicated along the axes in units of $M$) are adjusted to capture the orbits effectively. Orbits from ``more interesting" regions are purple, one chaotic orbit is green.}
\label{orbit-shapes}
\end{figure*}

\subsection{Resonance and chaos in orbit shapes}\label{shapes}

Poincar\'e surfaces of section represent a basic tool for assessing the overall structure of the possible test-particle motion, but one should keep in mind that they are really just {\em sections} through phase space, flattening out most of the information about individual trajectories. When comparing different systems, like the relativistic one and its pseudo-Newtonian counter-parts we are interested in here, one naturally first checks the Poincar\'e diagrams for analogies and variances, but in fact any statements concerning the occurrence of certain structures in Poincar\'e sections has to be taken with caution, because a particular sequence of recorded points (e.g. equatorial transitions) does not in general unveil a trajectory uniquely.

In order to get an idea of what trajectory shapes such structures may represent and to illustrate what the statements about the frequency ratios mean for the actual trajectories, we select one of the sections obtained within the series capturing the dependence on iMM1-disc mass, namely the ${\cal M}=0.35 M$ section of Fig. \ref{iMM1-mass-log} (where the black hole was simulated by the logarithmic potential). This case represents the weakest perturbation for which separatrix chaos already appears near the 3-fold island; the diagram is repeated in Fig. \ref{orbit-shapes-Poincare} with a selection of orbits plotted in colours. The motion in the $\phi$ direction is dynamically unimportant (bound by conserved integral $\ell$) in the axially symmetric case, so we suppress this dimension and illustrate the orbital shapes within the Weyl $(\rho,z)$ meridional plane. The results are grouped in Fig. \ref{orbit-shapes}, marked by the same colours as their equatorial sections in Fig. \ref{orbit-shapes-Poincare}.

There are two distinct structures in Fig. \ref{orbit-shapes-Poincare}, the 3-fold and the 5-fold island; the ratio of the radial to vertical frequencies is 2:3 for the former and 4:5 for the latter. The shapes of the trajectories reveal less thick resonances hidden in both the central and 3-fold island, but for most of them only a longer evolution track could confirm whether it is actually a resonance or a near-periodic orbit only. However, one can notice a certain deformation due to the proximity of a resonance in Fig. \ref{orbit-shapes}: the fish-like shape of the 2:3 trajectory corresponding to the 3-fold island is reflected in a significant part of the neighbouring non-resonant phase space, which might perhaps lead to observable signs in an ensemble of particles orbiting near the black hole. Besides obvious structures, one also notices, when recording data for Fig. \ref{orbit-shapes-Poincare}, that the computation of the single purple orbit lying within the (blue) regular single-periodic region takes much longer time than that of the other orbits around. This typically indicates that one is close to a resonance, which is confirmed in Fig. \ref{orbit-shapes}. Let us also point out to the rightmost orbit in the last-but-one row and to the middle one in the last row (both are purple): they are very similar, both lying just between a ``box" regime and a resonant regime of the regular region, and analogous to a rotation of a pendulum very close to libration; the spatial corridors more densely filled by the orbit in Fig. \ref{orbit-shapes-Poincare} then correspond to the pendulum near-stopping at the unstable top equilibrium (before falling back to rotation) which stands for an unstable counter-part of the stable periodic orbit at the resonance core.

As seen in the second row of Fig. \ref{orbit-shapes}, the time span corresponding to some 200 equatorial-plane intersections is not enough to discern between the regular trajectory (red) and the very close separatrix chaos (green). On the other hand, the respective surface of section in Fig. \ref{orbit-shapes-Poincare} allows to discern order and chaos unambiguously at the toll of 7400 equatorial intersections.
To better understand the computational/observational times required for a clear distinction between the regular and weakly chaotic orbit, we employ a time-series recurrence analysis in the following section.

\begin{figure*}
\includegraphics[width=\textwidth]{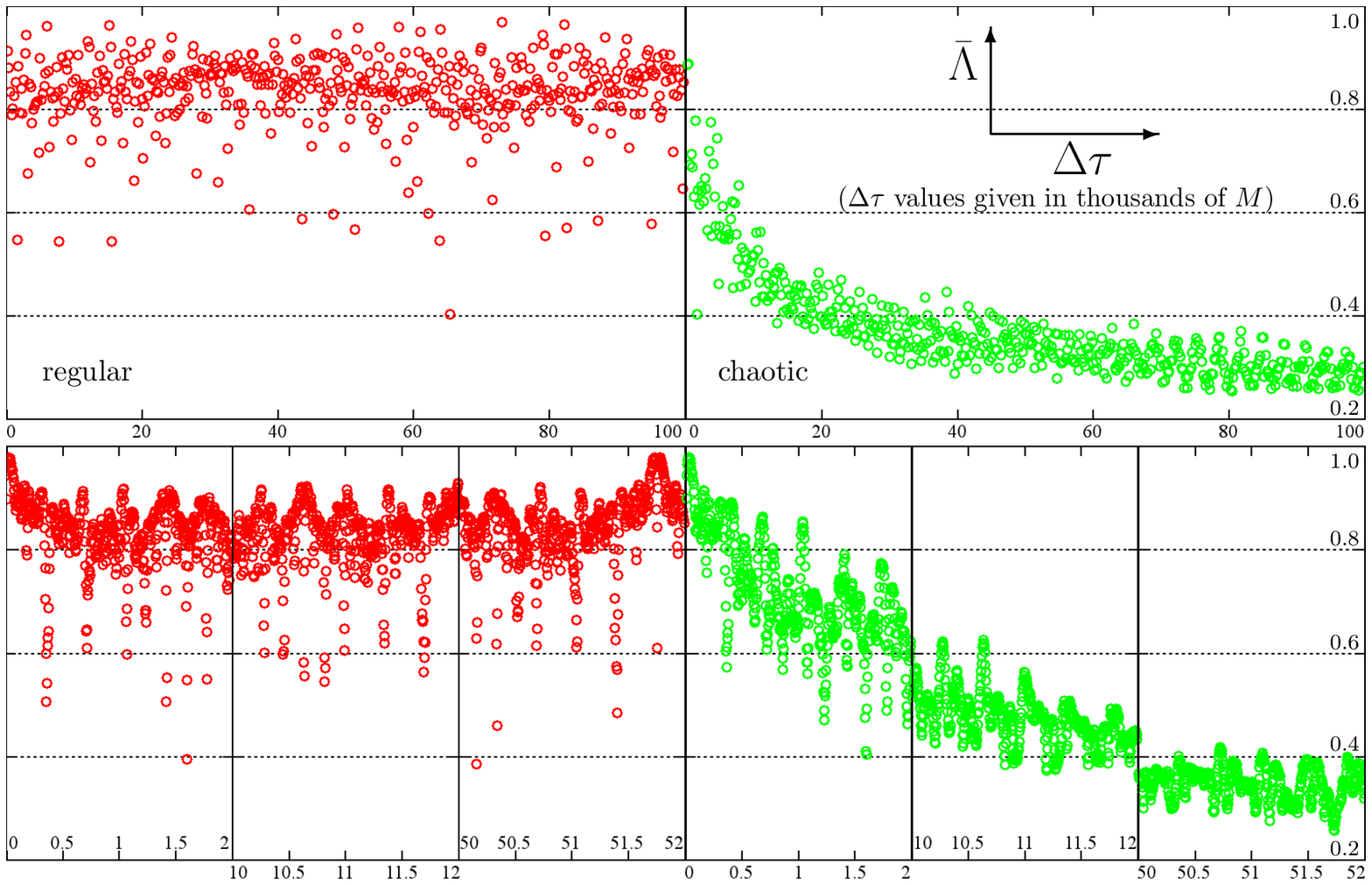}
\caption
{Two ``neighbouring" orbits from Fig. \ref{orbit-shapes-Poincare} (and \ref{orbit-shapes}), namely the outmost of the red-colour (3-fold) regular ones (left plot) and the green-colour chaotic one (right plot), are clearly distinguished by the Kaplan--Glass ``average directional vectors" recurrence method. The meaning of the $\bar{\Lambda}(\Delta\tau)$ dependence is explained in the main text. Recall that both orbits represent motion of free particles with ${\cal E}+1=0.955$, $\ell=3.75M$ in the field of a centre described by the logarithmic potential (mass $M$) and surrounded by the iMM1 disc with mass ${\cal M}=0.35M$ and inner radius $r_{\rm disc}=20M$. The orbits have been followed for about $500\,000M$ of proper time (some 5000 periods); the top row shows the dependence $\bar{\Lambda}(\Delta\tau)$ from $\Delta\tau=0$ up to $\Delta\tau=100\,000M$, while three selected intervals of $\Delta\tau$ are added in more detail in the bottom row (the $\Delta\tau$-axis labels are in thousands of $M$ everywhere).}
\label{WADV-2orbits}
\end{figure*}

\subsection{Recurrence analysis}

It is appropriate to support the Poincar\'e-section observations by some other independent method. Like in paper II, we turn to two recurrence methods here, one based on statistics over directions in which the orbits traverse a pre-selected mesh of phase-space cells, the other built on recurrences themselves to the neighbourhoods of phase-space points.

\cite{KaplanG-92} suggested to monitor the evolution of a tangent to the trajectory in small subsets of phase space which are crossed recurrently. For this purpose, the phase space is ``reconstructed" from a given data series $x(\tau)$ (either computed or measured) by adding the latter's replicas delayed by some shift $\Delta\tau$ and its multiples. The method was designed to distinguish between deterministic and random systems, but we saw in paper II that it is quite sensitive to weak irregularities and thus very well able to also recognize how chaotic the (deterministic) system is. Without going into details (see paper II for description of how we use the method for our system), let us only recall main points:
\begin{itemize}
\item
First the dimension $d$ is chosen of the phase space to be reconstructed, plus the delay $\Delta\tau$ and the size of boxes into which the phase space is divided.
\item
Average tangents of a trajectory within a given ($j$th) box are summed (vector addition) for a large number of recurrent transits and the length of the result is suitably normalized; the result is denoted as $V_j(\Delta\tau)$.
\item
The resulting norm is averaged then over all boxes which were crossed exactly $n$-times.
\item
The result depends on $n$, on $d$, on $\Delta\tau$ and on the lattice-box size. (The choice of these parameters in turn depends on how long data series one deals with.)
With $n$ it decreases roughly as $n^{-1/2}$ for random data, whereas more slowly for a deterministic system (in a theoretical limit of an infinitely long series and infinitesimally fine grain, it even remains 1 for the deterministic case).
The dependence on $\Delta\tau$ is specifically studied on the deviation of the result from the value obtained for random walk, computed for each box and then averaged over all occupied boxes,
\[\bar\Lambda=\bar\Lambda(\Delta\tau)
            :=\left\langle
              \frac{[V_j(\Delta\tau)]^2-(\bar{R}^d_{n_j})^2}{1-(\bar{R}^d_{n_j})^2}
              \right\rangle\]
($\bar{R}^d_{n_j}$ is the average displacement per step for random walk of length $n_j$ in $d$ dimensions).
In a theoretical limit, $\bar\Lambda=0$ for a random walk, whereas $\bar\Lambda=1$ for a deterministic system; in practice, $\bar\Lambda$ falls off roughly as autocorrelation function for a random series, while more slowly for a deterministic one.
\end{itemize}

We have subjected to the Kaplan--Glass test two orbits from Figs. \ref{orbit-shapes-Poincare} and \ref{orbit-shapes}, namely the outmost of the red-colour (3-fold) regular ones and the nearby green-colour chaotic one which has arised from a separatrix breakup. The autocorrelation corresponding to the dependence of the ``directional-vectors average" $\bar\Lambda$ on time delay $\Delta\tau$ clearly confirms the different character of the orbits. Let us specify that we started the analysis from reconstructing the phase space as three-dimensional and dividing it in $25^3$$=$$15625$ boxes; average number of transitions through one box (among those which were crossed at least once) has been around 50.

The second method rests on the statistics of recurrences to prescribed neighbourhoods of phase-space points (either of the ``original" phase space, or the reconstructed one). \cite{MarwanRTK-07} elaborated various useful outcomes of such a statistics and codified their computation in the {\sc recurrence plots} software; we already applied it, in paper II, to the exact relativistic system.

The main object of the analysis is the symmetric recurrence matrix
\begin{equation}
  R_{i,j}(\epsilon)=
  \Theta\left(\epsilon-\parallel\!\!{\boldsymbol X}_i-{\boldsymbol X}_j\!\!\parallel\right),
  \qquad i,j=1,...,N \;,
\end{equation}
where ${\boldsymbol X}_i={\boldsymbol X}(\tau_i)$ denote $N$ successive points of a given phase trajectory, $\epsilon$ is the radius of a chosen neighbourhood (called threshold), $\Theta$ is the Heaviside step function and $\parallel\cdot\parallel$ denotes the chosen norm (we use a simple Euclidean norm, but other can be considered, without significantly affecting the results). The matrix contains only units (meaning that $j$-th point is close to the $i$-th and represented by black dots) and zeros (blank positions which mean the opposite). For regular systems, the recurrences arrange in distinct structures, in particular in long parallel diagonal lines and checkerboard structures, whereas for random systems they are scattered without order; the chaotic systems provide something in between. A number of useful ``quantifiers" can then be extracted from the recurrence data, as explained in \cite{MarwanRTK-07} and also briefly reviewed in our paper II. The simplest ones follow from the lengths of the diagonal and vertical/horizontal lines which have occurred in the recurrence matrix.

The recurrence pattern clearly depends on the time step $\Delta\tau$ with which the trajectory is sampled and on the ``target" radius $\epsilon$. Besides that, the matrix often contains false recurrence records that should be discarded from statistics. For example, if $\epsilon$ is too large and the time step too small, several successive points of the trajectory {\em of course} lie within the $\epsilon$-neighbourhood of each other, but do not represent true recurrences.
Due to the same reason, the real recurrences may then involve more than one point, even if the orbit comes across its certain previous part in quite a divergent manner. To overcome such false signals, several further ``thresholds" are introduced and adjusted, mainly the minimal lengths of relevant diagonal and vertical lines, $l_{\rm min}$ and $v_{\rm min}$.

The choice of the recurrence threshold $\epsilon$ should also take into account the physical extent of the orbit and its variance. However, we overcome this ambiguity by rescaling the time series in such a way that each variable has zero mean and unit variance. This also assures that motion in all coordinate directions have equal weight in the analysis, irrespective of the actual ranges spanned by the trajectory.

\begin{figure*}
\includegraphics[width=\textwidth]{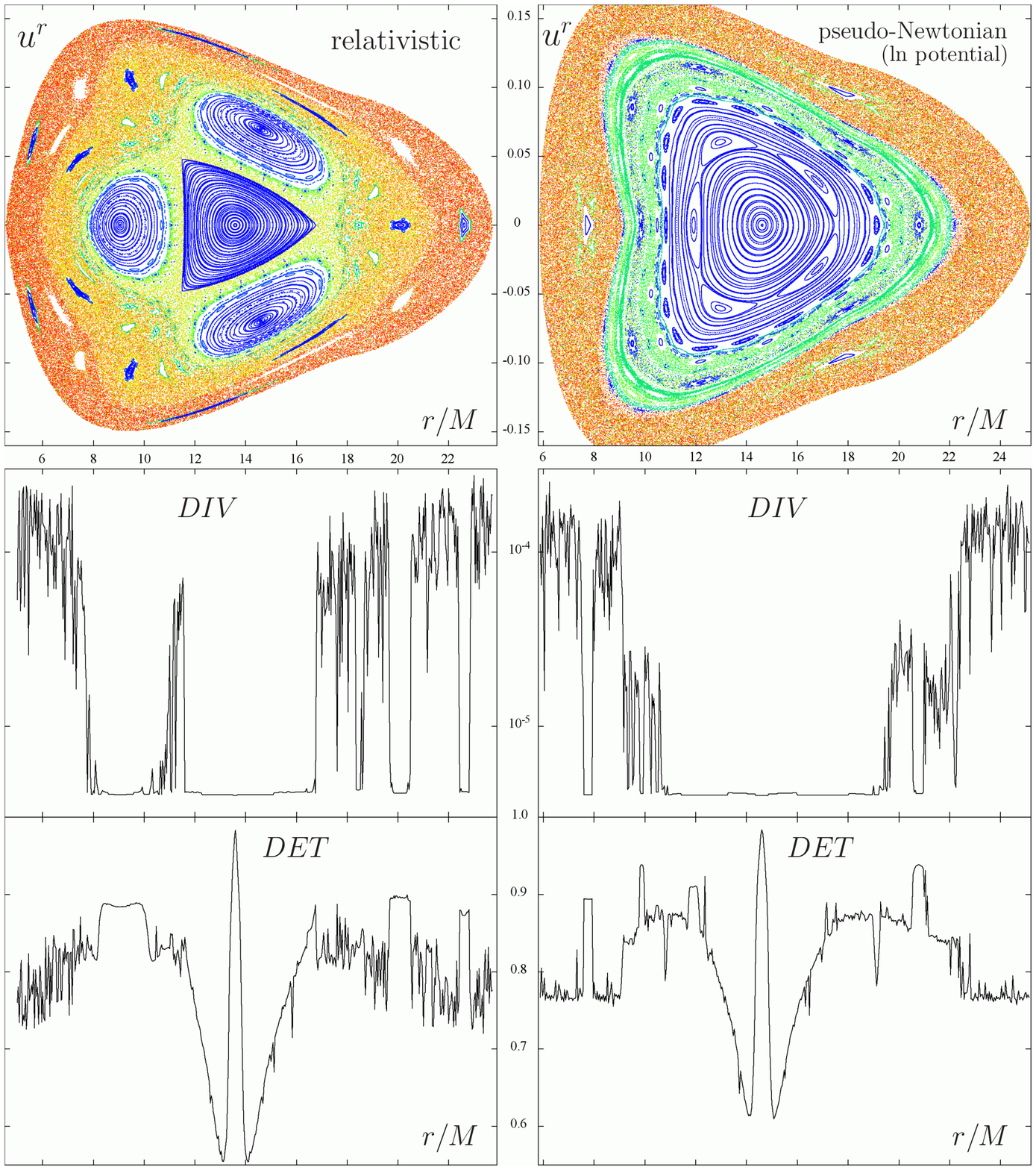}
\caption
{Examples of the recurrence-plot results, obtained for free motion with ${\cal E}(+1)=0.9532$ and $\ell=3.75M$ in the black-hole--disc field with ${\cal M}=0.5M$ and $r_{\rm disc}=18M$. Exact relativistic system is represented in the {\it left column}, while pseudo-Newtonian system employing the logarithmic potential is in the {\it right column}. The top row shows Poincar\'e diagrams coloured according to the value of $DIV$ whose $u^r=0$/$v^r=0$ radial profile is also plotted in the middle row (going from blue to red, the value of $DIV$ increases, which corresponds to increasing irregularity); the bottom row shows the same profile for another simple quantifier $DET$, given by ratio of the points which form a diagonal line longer than a certain minimum. The horizontal axes ($r$ in units of $M$) are common for all rows and the vertical axes are common for both columns. One more remark: notice that in the left Poincar\'e section the orange and red orbits are rather separated, whereas in the right one they are mixed within the chaotic sea. This is because the left section is actually divided into a ``sticky" interior region harbouring weaker chaos and only slowly diffusing particles, and the outer chaotic sea with strong chaos. In the right section, the outer layer is mixed orange-red, with its less chaotic orange trajectories perhaps corresponding to motion ``sticked", for a short time, to the three small islands on the outskirts. For ${\cal M}\approx 0.65M$ the green layer of very weak chaos gets connected with the outskirts, thus yielding a picture rather similar to the relativistic case.}
\label{12/II-counterpart}
\end{figure*}

For the relativistic--pseudo-Newtonian comparison using the recurrence-matrix analysis, we choose fig. 12 of paper II. There, several ``quantifiers" were computed for 470 geodesics having specific energy ${\cal E}=0.9532$ and specific angular momentum $\ell=3.75M$, sent tangentially (with $u^r=0$) from radii between $r=5M$ and $r=24M$ (with step $0.04M$) from the equatorial plane of the system of a black hole ($M$) and the iMM1 disc with ${\cal M}=0.5M$ and $r_{\rm disc}=18M$. The orbits were followed for about $250\,000M$ of proper time with ``sampling period" $\Delta\tau=45M$, the minimal length of diagonal/vertical lines has been set at $90M$ and the radius of the recurrence neighbourhood (the threshold) at $\epsilon=1.25$. Two of the quantifiers -- the most simple one called $DIV$, given by reciprocal of the longest recurrence-matrix diagonal, and a much more ``sophisticated" one read off from the slope of the histogram of diagonals (and providing an estimate of the maximal Lyapunov exponent), were particularly illustrated by colouring the computed orbits according to their values in the Poincar\'e diagram. Two main observations were made: i) all the quantifiers proved sensitive to even tiny phase-space features, and ii) the computationally easy $DIV$ quantifier proved equally efficient as the more sophisticated one.

The above recurrence analysis was performed in a 6D phase space ($r$, $\theta$, $\phi$ and the respective velocities), while, for the present comparison, we have repeated it, for the same set of geodesics, in the $(r\sin\theta,r\cos\theta)$ plane plus the respective velocity dimensions only. Elimination of $\phi$ from the analysis has some interesting aspects, even though it is a Killing-symmetry coordinate. For instance, note that in the full 3D configuration space there are virtually no true recurrences, since even the most regular central orbit is quasi-periodic in $\phi$. Within the meridional plane, on the other hand, the resonance cores produce true recurrences (see section \ref{shapes}).

Let us add that the $\phi$ coordinate can be viewed as a kind of ``dynamical memory", because
\begin{equation}
  \Delta\phi = \int_{\Delta t}\frac{\ell\;{\rm d}t}{r^2\sin^2\theta}
\end{equation}
(a relativistic formula only contains proper time $\tau$ instead of $t$). Hence, the inclusion of $\phi$ actually adds non-trivial information, so the change resulting from its elimination might indicate the robustness of various recurrence indicators.

The parameters we have used for the re-analysis of fig. 12 of paper II are $l_{\rm min}=3$ and $v_{\rm min}=3$ for the minimal diagonal and vertical, the Theiler window $w=3$ and the neighbourhood radius $\epsilon=0.8$ (with Euclidean metric used for the distance). The trajectories for the recurrence matrix were then recorded at a time step $\Delta\tau= 45M$ for a total time of about $250\,000M$ like in paper II.
As can be seen in Fig. \ref{12/II-counterpart}, the $DIV$ indicator is not changed by the 3D$\rightarrow$2D projection at all, whereas the $DET$ quantifier turned out to be less robust in this respect. In particular, the $DET$ quantifier seems to wrongly indicate that a large part of the central island is ``less deterministic" than the surrounding chaos. To understand this point, let us recall that the $DET$ indicator is defined as the ratio of the number of diagonal lines longer than $l_{\rm min}$ to the number of recurrence points. We checked that the orbits in the central island show a large number of recurrence points but not always grouped into longer diagonal lines. The performed normalization with respect to the total number of recurrence points thus has an undesirable effect in this case.

Now to the comparison: we take an analogous pseudo-Newtonian situation, namely the gravitational system with ``the same" parameters and with the central black hole simulated by the logarithmic potential (we do not employ the Paczy\'nski--Wiita one, because that yields rather open accessible region, which makes the chaotic sea efficiently drained away to the centre), and subject it to the same recurrence-matrix analysis as performed in fig. 12 of paper II; the results are given in Fig. \ref{12/II-counterpart}. Clearly the phase-space structure is rich for the given parameters and also rather different from its paper-II counter-part. (As already stressed above, the relativistic and pseudo-Newtonian systems are qualitatively similar, but the similar phase-space pictures are somewhat shifted with respect to each other in the parameter space.) The question is whether the corresponding recurrence patterns are still not alike, in spite of this first-sight difference.

In Fig. \ref{12/II-counterpart}, the left column is relativistic and the right column is pseudo-Newtonian (with the logarithmic potential), both plotted in the same scale. Both Poincar\'e sections are coloured by the longest-diagonal reciprocal $DIV$, the latter's zero-velocity radial profile being also plotted below, and the last row shows another simple quantifier $DET$, given by ratio of the points which form a diagonal line longer than a certain value within all the recurrence points. Although the surfaces of section reveal rather different structures, we do not see any big overall divergence in the recurrence characteristics.

\section*{Concluding remarks}

We have considered Newtonian dynamical systems describing the massive-test-particle motion in a gravitational field of a Schwarzschild-like centre simulated by a suitable potential and surrounded symmetrically by a gravitating thin disc or ring. Trying to learn how they differ from the corresponding relativistic system, namely the time-like geodesic dynamics in the field of a Schwarzschild black hole surrounded by ``the same" disc or ring (described by the same Newtonian potential), we plotted Poincar\'e diagrams of equatorial transitions for a number of similar situations (same coordinate position and relative mass of the disc or ring, same values of the particles' conserved energy and angular momentum) and found similar tendencies, typical for weakly non-integrable systems. The picture revealed by the surfaces of section was also confirmed by two recurrence methods, one resting on statistics over directions in which the orbits transit recurrently the boxes of a pre-selected phase-space mesh, and the other analysing the recurrences themselves to some prescribed neighbourhood of phase-space points. We have been using a {\em different} code than in previous papers of this series, so the present results also support robustness of the observations made.

A careful conclusion would be that the pseudo-Newtonian approach can reproduce the long-term dynamics of our relativistic system reasonably, though there appear various quantitative differences. However, this conclusion strongly depends on the potential used to mimic the black-hole centre: the Paczy\'nski--Wiita and the logarithmic potentials provide results very similar to the relativistic treatment, while the Nowak--Wagoner potential offers quite a different picture; some other potentials are not suitable for these purposes at all (although they may be efficient in another context). Yet even the Paczy\'nski--Wiita and the logarithmic potentials differ considerably (from the relativistic system as well as from each other) in the phase-space accessible region they determine -- the PW potential is too open, whereas the ln potential is too closed in direction toward the centre, which mainly affects how effectively the centre ``sucks out" the chaotic orbits; nevertheless, this does not seem to influence much the behaviour of regular structures under parameter change. Generally, the pseudo-systems (involving the PW and mainly the ln potential) can be labelled slightly more unstable than the exact relativistic system, since their phase-space structures evolve somewhat faster with parameters.

As mentioned in preceding papers, there are several possible directions of further study. One can certainly subject the dynamical system -- either the relativistic or the (pseudo-)Newtonian one -- to still other methods (than already employed there), e.g. the Melnikov-integral calculation or the basin-boundary analysis, or to a more detailed study of its particularly ``interesting" orbits (mainly the periodic ones). However, most important astrophysically is to make our setting more realistic and to try to confront it with what is going on in real celestial systems. The simplest issue, at least within the static and axially symmetric situation, would be to add another gravitating components like central star cluster, a jet or a halo. Second, we plan to consider non-singular (i.e. 3D) sources instead of the infinitesimally thin ones. This is especially important in systems where the orbits can reach very close to the sources, mainly in the relativistic description when the rings as well as edges of the thin discs usually represent a curvature singularity and (thus) the space-time is unnaturally deformed in their vicinity. In particular, one would be interested in how reasonably -- as far as the long-term dynamics is concerned -- the singular (Bach--Weyl) ring can approximate a toroidal source; this could be studied on a sequence of toroids gradually thinning to a ring.

Once obtaining a sufficiently realistic description, it is desirable to look for consequences for observational phenomena. For instance, the ensembles of initial conditions studied in the surfaces of section can be understood as {\em actual} collisionless ensembles of particles orbiting a black hole. The increased ``suck-in" of the ensemble under perturbation then imply an enhanced accretion rate, while the resonances, on the other hand, correspond to particularly behaving oscillation modes.

And then there are {\em difficult} aspects of ``realisticity": incorporating ({\em adequately}) rotation of the gravitating bodies (which brings dragging effects into the relativistic systems) and possibly also back reaction resulting from the non-test character of the orbiter.

\section*{Acknowledgements}

Access to computing and storage facilities owned by parties and projects contributing to the Czech National Grid Infrastructure MetaCentrum, provided under the programme ``Projects of Large Infrastructure for Research, Development, and Innovations" (LM2010005), is greatly appreciated.
The plots were produced with the help of the {\sc Gnuplot} utility, C. Louvrier's {\sc PNGslim} program and D. Krause's {\sc bmpp} program.
We are grateful for support from grants GAUK-2000314 (V.W.); GACR-14-10625S (O.S.); DEC-2012/05/E/ST9/03914 (P.S.).
O.S. also thanks M. Crosta for kind hospitality at the Osservatorio Astrofisico di Torino.


\begin{thebibliography}{99}

\bibitem[\protect\citeauthoryear{Artemova et al.}{1996}]
        {ArtemovaBN-96}
   Artemova I. V., Bj\"ornsson G., Novikov I. D., 1996,
   ApJ, 461, 565
\bibitem[\protect\citeauthoryear{Blanes \& Moan}{2002}]
        {BlanesM-02}
   Blanes S., Moan P. C., 2002,
   J. Comp. Appl. Math., 142, 313
\bibitem[\protect\citeauthoryear{Chakrabarti \& Mondal}{2006}]
        {ChakrabartiM-06}
   Chakrabarti S. K., Mondal S., 2006,
   MNRAS, 369, 976
\bibitem[\protect\citeauthoryear{Crispino et al.}{2011}]
        {Crispino-etal-11}
   Crispino L. C. B., da Cruz Filho J. L. C., Letelier P. S., 2011,
   Phys. Lett. B, 697, 506
\bibitem[\protect\citeauthoryear{D'Afonseca et al.}{2005}]
        {DAfonsecaLO-05}
   D'Afonseca L. A., Letelier P. S., Oliveira S. R., 2005,
   Class. Quantum Grav., 22, 3803
\bibitem[\protect\citeauthoryear{Ghosh et al.}{2014}]
        {GhoshSB-14}
   Ghosh S., Sarkar T., Bhadra A., 2014,
   MNRAS, 445, 4463
\bibitem[\protect\citeauthoryear{Gu\'eron \& Letelier}{2001}]
        {GueronL-01}
   Gu\'eron E., Letelier P. S., 2001,
   Astron. Astrophys., 368, 716
\bibitem[\protect\citeauthoryear{Hairer et al.}{2006}]
        {HairerWL-06}
   Hairer E., Wanner G., Lubich C.,  2006,
   Geometric Numerical Integration. Structure-Preserving Algorithms for Ordinary Differential Equations,
   Springer Ser. in Comp. Math. 31, Springer, Berlin Heidelberg
\bibitem[\protect\citeauthoryear{Ivanov \& Prodanov}{2005}]
        {IvanovP-05}
   Ivanov R. I., Prodanov E. M., 2005,
   Phys. Lett. B, 611, 34
\bibitem[\protect\citeauthoryear{Kaplan \& Glass}{1992}]
        {KaplanG-92}
   Kaplan D. T., Glass L., 1992,
   Phys. Rev. Lett., 68, 427
\bibitem[\protect\citeauthoryear{Karas \& Abramowicz}{2015}]
        {KarasA-15}
   Karas V., Abramowicz M. A., 2015,
   in Proc. RAGtime 14--16: Workshop on black holes and neutron stars,
   eds. S. Hled\'{\i}k, Z. Stuchl\'{\i}k
   (Silesian Univ. in Opava, Opava 2015), to appear
   (arXiv:1412.6832 [astro-ph.HE])
\bibitem[\protect\citeauthoryear{Marwan et al.}{2007}]
        {MarwanRTK-07}
   Marwan N., Romano M. C., Thiel M., Kurths J., 2007,
   Phys. Reports, 438, 237
\bibitem[\protect\citeauthoryear{Nowak \& Wagoner}{1991}]
        {NowakW-91}
   Nowak M. A., Wagoner R. V., 1991,
   ApJ, 378, 656
\bibitem[\protect\citeauthoryear{Paczy\'nski \& Wiita}{1980}]
        {PaczynskiW-80}
   Paczy\'nski B., Wiita P. J., 1980,
   A\&A, 88, 23
\bibitem[\protect\citeauthoryear{\c{S}elaru et al.}{2005}]
        {SelaruMCG-05}
   \c{S}elaru D., Mioc V., Cucu-Dumitrescu C., Ghenescu M., 2005,
   Astron. Nachr., 326, 356
\bibitem[\protect\citeauthoryear{Semer\'ak}{2015}]
        {Semerak-15}
   Semer\'ak O., 2015,
   ApJ, 800, 77
\bibitem[\protect\citeauthoryear{Semer\'ak \& Karas}{1999}]
        {SemerakK-99}
   Semer\'ak O., Karas V., 1999,
   A\&A, 343, 325
\bibitem[\protect\citeauthoryear{Semer\'ak \& Sukov\'a}{2010}]
        {SemerakS-10}
   Semer\'ak O., Sukov\'a P., 2010,
   MNRAS, 404, 545 (paper I)
\bibitem[\protect\citeauthoryear{Semer\'ak \& Sukov\'a}{2012}]
        {SemerakS-12}
   Semer\'ak O., Sukov\'a P., 2012,
   MNRAS, 425, 2455 (paper II)
\bibitem[\protect\citeauthoryear{Semer\'ak et al.}{1999}]
        {SemerakZZ-99b}
   Semer\'ak O., \v{Z}\'a\v{c}ek M., Zellerin T., 1999,
   MNRAS, 308, 705
\bibitem[\protect\citeauthoryear{Seyrich \& Lukes-Gerakopoulos}{2012}]
        {SeyrichLG-12}
   Seyrich J., Lukes-Gerakopoulos G., 2012,
   Phys. Rev. D, 86, 124013
\bibitem[\protect\citeauthoryear{Steklain \& Letelier}{2006}]
        {SteklainL-06}
   Steklain A. F., Letelier P. S., 2006,
   Phys. Lett. A, 352, 398
\bibitem[\protect\citeauthoryear{Steklain \& Letelier}{2009}]
        {SteklainL-09}
   Steklain A. F., Letelier P. S., 2009,
   Phys. Lett. A, 373, 188
\bibitem[\protect\citeauthoryear{Stoffer}{1995}]
        {Stoffer-95}
   Stoffer D., 1995,
   Computing, 55, 1
\bibitem[\protect\citeauthoryear{Stuchl\'{\i}k \& Kov\'a\v{r}}{2008}]
        {StuchlikK-08}
   Stuchl\'{\i}k Z., Kov\'a\v{r} J., 2008,
   Int. J. Mod. Phys. D, 17, 2089
\bibitem[\protect\citeauthoryear{Sukov\'a \& Semer\'ak}{2013}]
        {SukovaS-13}
   Sukov\'a P., Semer\'ak O., 2013,
   MNRAS, 436, 978 (paper III)
\bibitem[\protect\citeauthoryear{Tejeda \& Rosswog}{2013}]
        {TejedaR-13}
   Tejeda E., Rosswog S., 2013,
   MNRAS, 433, 1930
\bibitem[\protect\citeauthoryear{Tejeda \& Rosswog}{2014}]
        {TejedaR-14}
   Tejeda E., Rosswog S., 2014,
   Generalized Newtonian description of particle motion in spherically symmetric spacetimes,
   submitted (arXiv:1402.1171 [gr-qc])
\bibitem[\protect\citeauthoryear{Vokrouhlick\'y \& Karas}{1998}]
        {VokrouhlickyK-98}
   Vokrouhlick\'y D., Karas V., 1998,
   MNRAS, 298, 53
\bibitem[\protect\citeauthoryear{Wang \& Wu}{2011}]
        {WangW-11}
   Wang Y., Wu X., 2011,
   Commun. Theor. Phys., 56, 1045
\bibitem[\protect\citeauthoryear{Wang \& Wu}{2012}]
        {WangW-12}
   Wang Y., Wu X., 2012,
   Chin. Phys. B, 21, 050504
\bibitem[\protect\citeauthoryear{Wang et al.}{2013}]
        {WangWS-13}
   Wang Y., Wu X., Sun W., 2013,
   Commun. Theor. Phys., 60, 433
\bibitem[\protect\citeauthoryear{Wegg}{2012}]
        {Wegg-12}
   Wegg C., 2012,
   ApJ, 749, 183
\bibitem[\protect\citeauthoryear{Yoshida}{1990}]
        {Yoshida-90}
   Yoshida H., 1990,
   Phys. Lett. A, 150, 262
\bibitem[\protect\citeauthoryear{\v{Z}\'a\v{c}ek \& Semer\'ak}{2002}]
        {ZacekS-02}
   \v{Z}\'a\v{c}ek M., Semer\'ak O., 2002,
   Czechosl. J. Phys., 52, 19

\end{thebibliography}
\end{document}